\documentclass[conference]{IEEEtran}
\IEEEoverridecommandlockouts

\usepackage{cite}
\usepackage{amsmath,amssymb,amsfonts}
\usepackage{algorithm}  
\usepackage{algorithmic}
\usepackage{subcaption}  
\usepackage{caption}     
\usepackage{float}       
\usepackage{textcomp}
\usepackage[table]{xcolor}
\usepackage{multirow}
\usepackage{graphicx}\graphicspath{{figures/}}
\usepackage{booktabs}     
\usepackage{array}        
\usepackage{siunitx}      
\usepackage{paralist}

\usepackage{hyperref}

\usepackage{tabularx}

\usepackage{flushend}

\newcommand{\sstitle}[1]{\smallskip\noindent\textbf{#1.} }

\def\Snospace~{\S{}}

\begin{document}

\setlength{\belowdisplayskip}{1pt}
\setlength{\belowdisplayshortskip}{1pt}
\setlength{\abovedisplayskip}{1pt}
\setlength{\abovedisplayshortskip}{1pt}

\title{An Efficient and Effective Agentic Group Shilling Attack on Recommender Systems}



\author{%
\IEEEauthorblockN{%
Quoc Viet Nguyen\textsuperscript{1},\quad
Trinh Pham\textsuperscript{1},\quad
Viet Huynh\textsuperscript{2},\quad
Hongzhi Yin\textsuperscript{3,*},
Quoc Viet Hung Nguyen\textsuperscript{1,*},
\\
Bay Vo\textsuperscript{4},\quad
Thanh Tam Nguyen\textsuperscript{1}%
}
\IEEEauthorblockA{%
\textsuperscript{1}\textit{Griffith University}, Australia\\
\textsuperscript{2}\textit{Edith Cowan University}, Australia\\
\textsuperscript{3}\textit{The University of Queensland}, Australia\\
\textsuperscript{4}\textit{Faculty of Information Technology, HUTECH University}, Vietnam%
}%
\thanks{\textsuperscript{*}Corresponding authors.}%
}

\maketitle

\begin{abstract}
Recommender systems have become core infrastructure for modern online platforms, personalizing content at scale and strongly influencing what users see, click on, and purchase. However, this dependence on user interaction also exposes them to shilling attacks, where malicious actors can inject fake profiles to distort item rankings and control visibility. Existing attacks often rely on target-specific fine-tuning or fixed profile templates, making them either difficult to adapt to different victims or easier to detect. To overcome these limitations, we propose the Agentic Group Attack System (AGAS), a coordinated shilling framework where a central Coordinator directs a group of role-switching worker agents to adaptively promote a target item across different victim families. The Coordinator dynamically adjusts the strategy when progress stalls or suppression signals increase, while workers pursue a shared objective and switch between active and inactive roles to avoid repetitive patterns. Under the same attack budgets and evaluation protocols, AGAS consistently surpasses strong baselines in target promotion while better preserving benign recommendation quality, weakening representative detectors, and achieving higher efficiency than prior attacks. These findings also emphasize that defending recommender systems may require mechanisms that can handle adaptive shilling campaigns, not just isolated fake-profile injections. Our code is available at \url{https://github.com/phkhanhtrinh23/AGAS}.

\end{abstract}

\begin{IEEEkeywords}
Recommender Systems, Group Shilling Attacks, Agentic AI.
\end{IEEEkeywords}

\section{Introduction}
\label{sec:introduction}

\begin{figure}[t]
    \centering
    \includegraphics[width=0.85\linewidth]{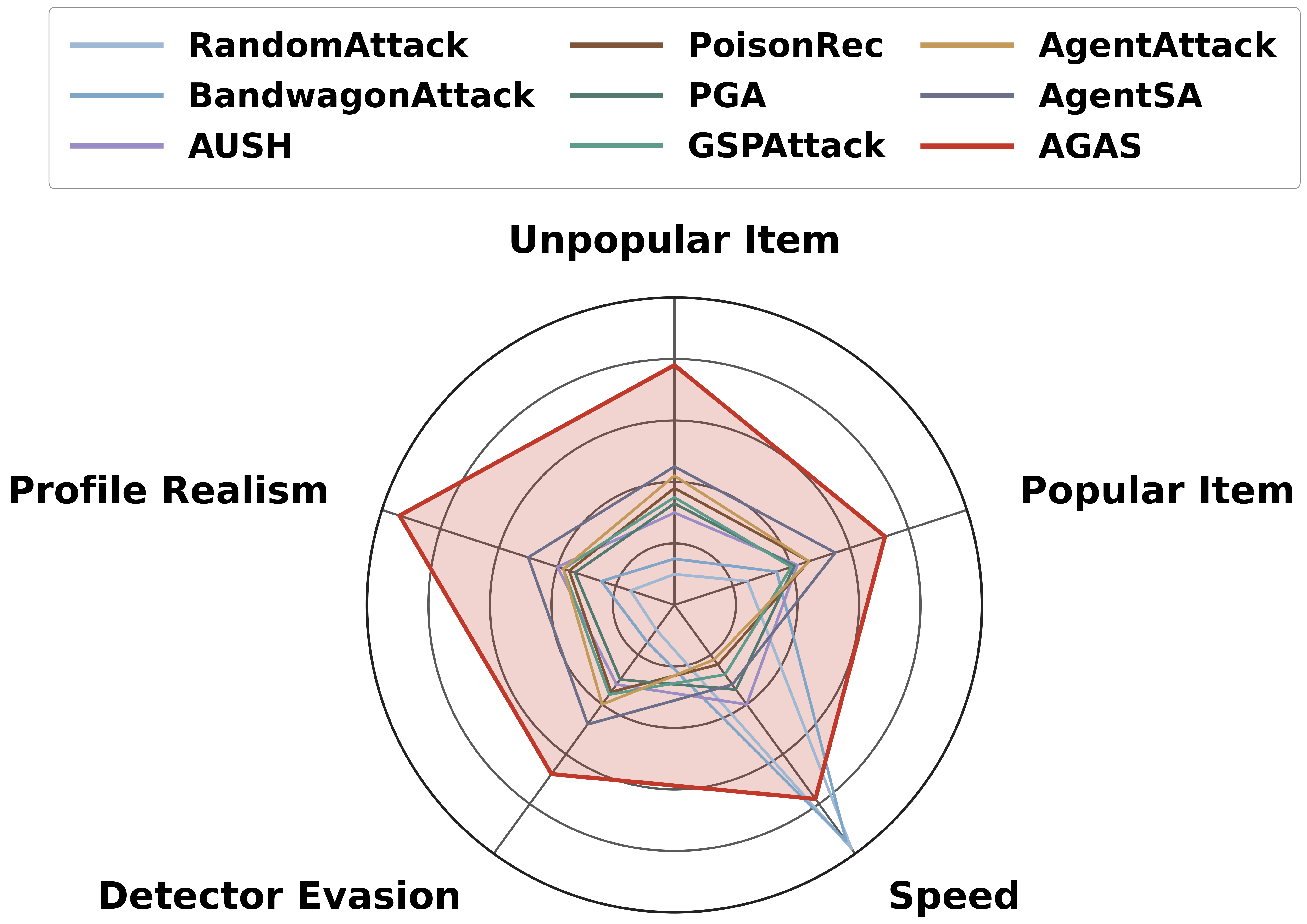}
    \caption{AGAS vs prior attacks on five axes.}
    \label{fig:intro_radar_motivation}
    \vspace{-1em}
\end{figure}

Recommender systems (RecSys) are widely deployed to rank products, media, and information at scale. In practice, many RecSys rely on Collaborative Filtering (CF), which learns from user--item interaction signals such as clicks, ratings, and purchases~\cite{koren2009matrix_factorization,he2017neumf,wang2019ngcf,he2020lightgcn}. This interaction-driven design also makes CF-based victims vulnerable to shilling attacks, where adversaries inject fake profiles to promote chosen target items. Prior attacks include heuristic profile injection~\cite{omahony2005attack_types}, augmented profile generation~\cite{lin2020aush,WU2021683}, surrogate-guided and bi-level optimization~\cite{li2016pga_cf_poisoning,nguyen2023gspattack,wang2024unveiling_contrastive_poisoning,zhao2025dual_promotion_poisoning}, reinforcement learning~\cite{fang2020poisonrec,fan2021copyattack}, and recent LLM-driven shilling attacks~\cite{gu2026llm_agent_shilling_wsdm,li2026agentattack,ning2024cheatagent,yang2025drunkagent,wang2024multi}. Despite this diversity,~\autoref{fig:intro_radar_motivation} shows that these methods are still limited in many aspects. The learning-based pipelines (AUSH~\cite{lin2020aush}, GOAT~\cite{WU2021683}, GSPAttack~\cite{nguyen2023gspattack}, and CLeaR~\cite{wang2024unveiling_contrastive_poisoning}) require extensive offline fine-tuning before a single campaign can be launched: AUSH must fit a GAN~\cite{goodfellow2020generative} on the clean interaction matrix for every new target. GOAT must train a graph-convolutional GAN to approximate the real rating distribution, introducing substantial preparation overhead before attack generation. GSPAttack and CLeaR must converge a bi-level optimization loop that alternates inner and outer updates. AgentSA~\cite{gu2026llm_agent_shilling_wsdm} avoids offline fine-tuning by using LLM agents to generate fake profiles, but each profile is still produced largely as an independent attacker following a shared prompt recipe, which can leave repeated behavioral patterns. AgentAttack~\cite{li2026agentattack} introduces an LLM planner, but the planner operates over a library of pre-generated profiles from existing attack families~\cite{lam2004shilling,omahony2005attack_types,WU2021683,lin2020aush}. The planner can only sample these limited candidates even when the pre-generated profiles share similar templates or fail to match a new victim setting. Most existing attacks operate as collections of independent fake users rather than coordinated \textit{group} campaigns.

The emergence of the Agentic Web~\cite{yang2025agentic,bernerslee2025this_is_for_everyone} changes the threat surface. Websites increasingly grant AI agents broad permissions to rate items, post reviews, and make purchases on users' behalf. Malicious actors can recruit a group of such agents and coordinate them to attack a RecSys~\cite{AIAgentTraps2026}. To the best of our knowledge, previous works have not explicitly studied coordinated agentic group shilling attacks in RecSys. They share some structural shortcomings. \textit{(1)~Costly preparation}: learning-based methods require hours of training before the first fake profile can be produced, and LLM-driven methods still generate expensive fake profiles or recursively sample new batches when the attack underperforms. \textit{(2)~Static patterns}: every fake user is locked to one generation algorithm or profile template for the entire campaign, leaving detectable distributional fingerprints. \textit{(3)~Static strategy}: there is no mechanism to redirect the group when the campaign underperforms. For example, fake users may rate the same suboptimal sequence of items.

This motivates us to propose the \textbf{Agentic Group Attack System (AGAS)}. Powered by the growing planning and reasoning capabilities of LLMs~\cite{wang2024llm_agent_survey,pham2025multilingual,fusionsql,pham2026learning,nguyen2026survey,Le_Pham_Quan_Luu_2024,pham-etal-2024-unibridge}, AGAS addresses each shortcoming in turn. \textit{(1)~Training-free}: A hidden Coordinator autonomously orchestrates the group from victim feedback, worker signals, and safety checks, without any training costs. \textit{(2)~Coordination}: the Coordinator jointly decides which role a worker should take, and when the group should attack, slow down, or stay inactive. This makes the campaign adaptive at the group level, rather than a set of independent fake users repeatedly pushing the same target. \textit{(3)~Role-switching}: Each worker is a stateful agent with its own memory. It switches roles (Profiler, Sniper, Camouflageur, or Inactive) across rounds under the Coordinator's supervision, reliably breaking the stable patterns that detectors rely on. \textit{(4)~Dynamic strategies}: AGAS switches strategies (\autoref{sec:strategy}) based on feedback from victims. This allows it to push aggressively when safe, slow down when risky, and avoid detectable attack patterns. This design lifts all five axes in~\autoref{fig:intro_radar_motivation}. It improves Speed by eliminating training latency, and improves Detector Evasion, Profile Realism, and Unpopular and Popular Item scores through coordination, role-switching, and dynamic strategies. Our contributions are summarized as follows:
\begin{compactitem}
\item \emph{Formulation}: To the best of our knowledge, this is the first formulation of coordinated group shilling as an adaptive campaign, where multiple fake users are jointly managed to maximize target exposure while maintaining stealth.
\item \emph{Method}: We propose \textbf{AGAS}, an agentic framework where a Coordinator manages specialized worker roles instead of treating fake users as independent attackers. It uses feedback signals to adjust role assignments and attack strategies without offline fine-tuning.
\item \emph{Evaluation}: Across different datasets and victim models, AGAS consistently outperforms conventional and agentic baselines under the same evaluation protocols, while preserving benign recommendation quality and weakening representative detectors.
\item \emph{Benchmark}: AGAS is both effective and efficient since it enables fast stress-testing of RecSys against sophisticated shilling attacks without costly fine-tuning or repeated surrogate retraining.
\end{compactitem}

\begin{figure*}[t]
    \centering
    \includegraphics[width=0.9\textwidth]{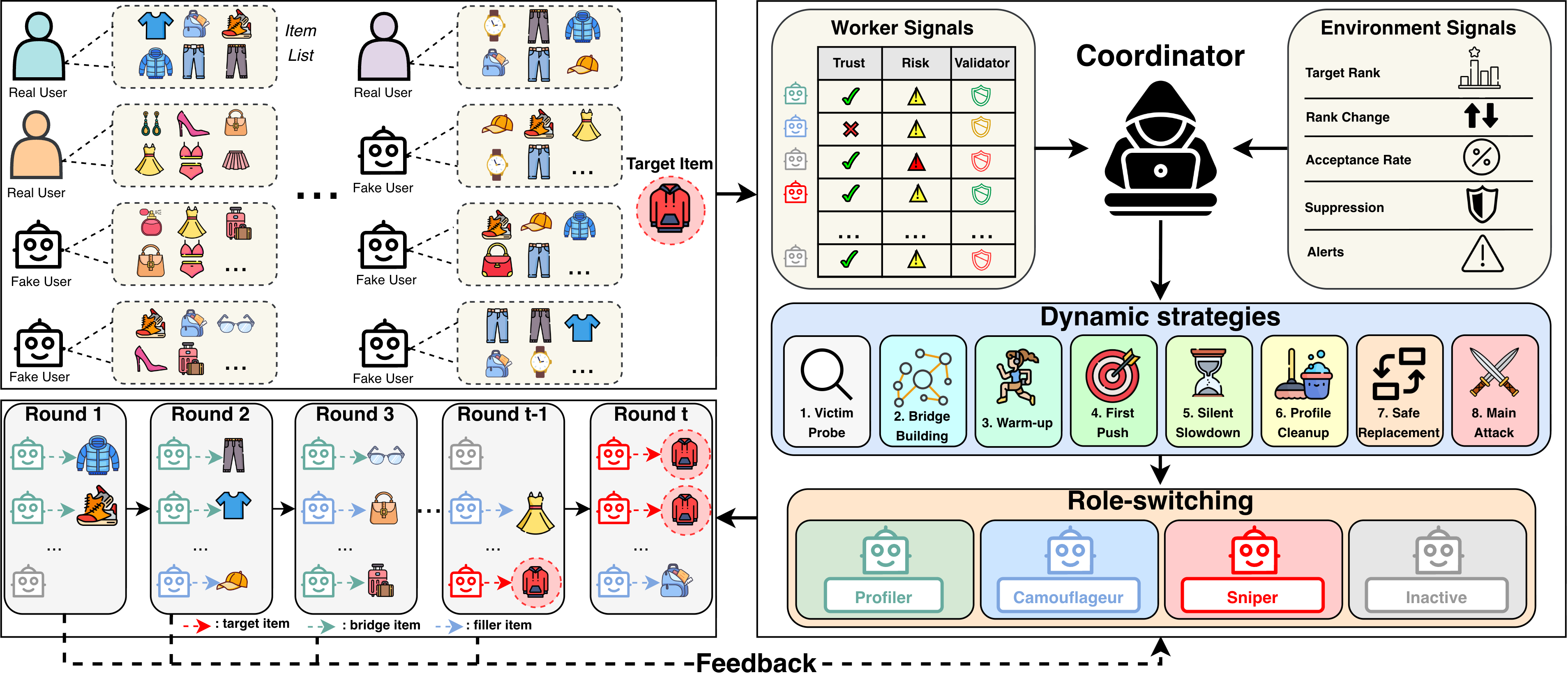}
    \caption{\textbf{Overview of the AGAS pipeline.} Across attack rounds, the Coordinator aggregates worker and environment signals to adapt fake-user roles and attack strategies, progressively steering the campaign toward effective promotion of the target item. AGAS effectively uses feedback-derived signals to probe the victim, slow down when suppression emerges, and keep some workers inactive to avoid repetitive or overly synchronized behavior.}
    \label{fig:agas_pipeline}
    \vspace{-1em}
\end{figure*}

\section{Related Work}

\label{sec:related}

Model-based CF learns user and item representations from historical interactions and remains widely used in RecSys~\cite{ricci2011recsys_handbook}. We consider two main families. \emph{Embedding-based} models include MF~\cite{koren2009matrix_factorization}, NMF~\cite{lee2000nmf}, NCF, NeuMF, and GMF~\cite{he2017neumf}. \emph{Graph-based} models propagate embeddings over user--item graphs, including NGCF~\cite{wang2019ngcf}, LightGCN~\cite{he2020lightgcn}, SimGCL~\cite{yu2022simgcl}, XSimGCL~\cite{yu2023xsimgcl}, EGCF~\cite{zhang2024egcf}, and LightCCF~\cite{zhang2025lightccf}. Existing attacks are often tailored to specific victim families, while AGAS operates in a black-box setting and adapts from victim feedback that is observable to the attacker.

\sstitle{White-box attacks}
White-box attacks assume access to victim objectives, gradients, or parameters. PGA~\cite{li2016pga_cf_poisoning} derives poisoning gradients for factorization models, while CLeaR~\cite{wang2024unveiling_contrastive_poisoning} and InfoAtk~\cite{ma2024stealthy} target contrastive or graph-based recommenders. Such access can enable strong attacks but limits practical applicability.

\sstitle{Gray and Black-box attacks}
Gray-box attacks use partial knowledge, typically through surrogate models. AIA~\cite{zhang2024adversarial_injection} learns transferable fake-user behaviors, while GSPAttack~\cite{nguyen2023gspattack} generates fake users and graph edges for GNN recommenders. Their effectiveness still depends on surrogate quality and victim-family assumptions. Black-box attacks use only observable behavior. Random Attack~\cite{lam2004shilling}, Bandwagon Attack~\cite{omahony2005attack_types}, HybridAttack~\cite{zhang2011hybrid}, and its social extension~\cite{yu2017hybrid_attacks} use fixed heuristics or templates. AUSH~\cite{lin2020aush} and GOAT~\cite{WU2021683} learn profile generators, while PoisonRec~\cite{fang2020poisonrec}, CopyAttack~\cite{fan2021copyattack}, and MultiAttack~\cite{wang2024multi} use reinforcement learning. R-Trojan~\cite{zhang2023incorporated} additionally exploits textual reviews. These methods improve profile generation but often require costly generative or policy training before deployment.

\sstitle{Agentic attacks}
Recent LLM-based methods introduce agents into black-box recommender attacks. CheatAgent~\cite{ning2024cheatagent} uses a single LLM agent against LLM-empowered recommenders. AgentSA~\cite{gu2026llm_agent_shilling_wsdm} assigns one agent to each fake account and rewrites profiles until validation succeeds. AgentAttack~\cite{li2026agentattack} uses an LLM planner to select pre-generated attack profiles from existing methods~\cite{lam2004shilling,omahony2005attack_types,lin2020aush,WU2021683}, with weak combinations filtered through surrogate retraining. In contrast, AGAS performs adaptive campaign management. A Coordinator manages role-switching workers, observes round-level feedback, pauses or slows attacks under suppression, replaces risky workers, and updates roles and strategies. Therefore, the attack evolves across rounds without relying on fixed profiles, prompts, or pre-generated libraries.

\section{Formulation}
\label{sec:formulation}

\subsection{Collaborative Filtering with Implicit Feedback}
\label{sec:formulation_cf}

Let $\mathcal{U}=\{1,\dots,M\}$ and $\mathcal{I}=\{1,\dots,N\}$ denote the user and item sets. The implicit matrix $\mathbf{R}\in\{0,1\}^{M\times N}$ has $R_{ui}=1$ if user $u$ interacted with item $i$, and $R_{ui}=0$ otherwise. CF learns a scoring function $\hat{y}_{ui}=f_{\Theta}(u,i)$ from each user's positive set $\mathcal{I}_u^{+}=\{i\in\mathcal{I}\mid R_{ui}=1\}$. This formulation covers common CF backbones, including matrix factorization~\cite{he2017neumf}, graph convolution~\cite{he2020lightgcn}, and graph-contrastive recommenders~\cite{yu2023xsimgcl}. BPR~\cite{rendle2009bpr} is used as the training objective:
\begin{equation}
\mathcal{L}_{\mathrm{BPR}}(\Theta)=
\sum_{(u,i,j)\in\mathcal{D}}
-\log \sigma(\hat{y}_{ui}-\hat{y}_{uj})
+\lambda \lVert \Theta\rVert_2^2 ,
\label{eq:bpr}
\end{equation}
where $\mathcal{D}=\{(u,i,j)\mid u\in\mathcal{U}, i\in\mathcal{I}_u^{+}, j\notin\mathcal{I}_u^{+}\}$, $\sigma(\cdot)$ is the sigmoid, and $\lambda$ is the regularization weight. At inference, the recommender returns:
\begin{equation}
\mathrm{TopK}(u)=
\operatorname*{arg\,topK}_{q\in \mathcal{I}\setminus \mathcal{I}_u^{+}}
\hat{y}_{uq}.
\label{eq:topk_def}
\end{equation}

\subsection{Targeted Shilling Attack}
The attacker selects a target item $i_{\mathrm{tar}}\in\mathcal{I}$ and aims to push it into the top-$K$ list of benign test users $\mathcal{U}_{\mathrm{test}}$, measured by:
\begin{equation}
\mathrm{HR}@K(i_{\mathrm{tar}})=
\frac{1}{|\mathcal{U}_{\mathrm{test}}|}
\sum_{u\in\mathcal{U}_{\mathrm{test}}}
\mathbb{I}[i_{\mathrm{tar}}\in\mathrm{TopK}(u)] .
\label{eq:hr_def}
\end{equation}
Following standard targeted shilling evaluation, the attacker injects fake users as new accounts, while attack success is measured on held-out benign users by checking whether the target item appears in their top-$K$ recommendation lists.

\sstitle{Threat model}
We assume a black-box setting: the attacker has no access to deployed-victim parameters, gradients, training procedure, evaluation-user identities, held-out interactions, or exact platform discount weights. The attacker injects $M_f$ fake users $\mathcal{U}_f=\{1,\dots,M_f\}$, where each fake user $w$ has interaction vector $\tilde{\mathbf{r}}_w\in\{0,1\}^{N}$. Stacking them gives $\tilde{\mathbf{R}}\in\{0,1\}^{M_f\times N}$, appended to the clean data:
\begin{equation}
\mathbf{R}'=\bigl[\mathbf{R};\,\tilde{\mathbf{R}}\bigr].
\label{eq:poison_append}
\end{equation}


\sstitle{Attack objective}
Let $\mathcal{A}(\cdot)$ be the training procedure mapping a poisoned matrix to a deployed recommender. AGAS maximizes target exposure while keeping injected profiles within a stealth tolerance:
\begin{equation}
\begin{aligned}
\max_{\{\tilde{\mathbf{r}}_w\}_{w\in\mathcal{U}_f}}\ &
\frac{1}{|\mathcal{U}_{\mathrm{test}}|}
\sum_{u\in\mathcal{U}_{\mathrm{test}}}
\mathbb{I}\!\left[i_{\mathrm{tar}}\in\mathrm{TopK}\!\left(u;\mathcal{A}(\mathbf{R}')\right)\right]\\
\text{s.t.}\ \ & \mathcal{S}(\tilde{\mathbf{R}};\mathbf{R})\le\varepsilon ,
\end{aligned}
\label{eq:attack_problem}
\end{equation}
where $\mathrm{TopK}(u;\mathcal{A}(\mathbf{R}'))$ is the top-$K$ list returned to benign test user $u$ by the recommender trained on the poisoned matrix $\mathbf{R}'$, $\mathcal{S}(\cdot;\cdot)$ measures anomaly relative to benign profiles, and $\varepsilon$ is the attacker--defender tolerance. In AGAS, this tolerance is operationalized through the worker and environment signals in \autoref{sec:method}, which decide whether to continue, slow down, pause, or switch strategy. AGAS constructs $\tilde{\mathbf{R}}$ over $T$ rounds using only attacker-accessible feedback, obtained from recommendation lists returned to its own fake accounts rather than from $\mathcal{U}_{\mathrm{test}}$ or any hidden metric of the deployed victim (\autoref{sec:method}). The reported attack uses the full poisoning history accumulated over all $T$ rounds, so $\mathbf{R}'$ in \autoref{eq:poison_append} is the matrix on which we report HR@\(\!K\) and NDCG@\(\!K\) in \autoref{sec:experiments}.

\section{Method}
\label{sec:method}

We present the \textbf{Agentic Group Attack System (AGAS)}, as shown in \autoref{fig:agas_pipeline}, where a Coordinator manages a pool of fake users over multiple rounds. The team contains four worker roles: \emph{Profiler} (PR), \emph{Camouflageur} (CA), \emph{Sniper} (SN), and \emph{Inactive} (IN). The Coordinator tracks global attack progress, worker safety, and campaign state, then adjusts both worker roles and attack strategy over time, without training latency. The complete process is described in \autoref{alg:agas_end_to_end}.

\begin{algorithm}[t]
\caption{AGAS Attack Procedure}
\label{alg:agas_end_to_end}
\resizebox{0.95\columnwidth}{!}{%
\begin{minipage}{\columnwidth}
\begin{algorithmic}[1]
\STATE \textbf{Input:} clean matrix $\mathbf{R}$, target item $i_{\mathrm{tar}}$, fake users $\mathcal{U}_f$, per-user budget $L$, rounds $T$.
\STATE \textbf{Output:} poisoned matrix $\mathbf{R}'=[\mathbf{R};\,\tilde{\mathbf{R}}]$.
\STATE Initialize Coordinator memory, worker states, victim family as unknown, and $\tilde{\mathbf{R}}\leftarrow\varnothing$.
\STATE Initialize worker signals as $\tau_{0,w}=0$, $\gamma_{0,w}=0$, and $\phi_{0,w}=0$ for each $w\in\mathcal{U}_f$.
\STATE Initialize environment signals as $\eta_0=1$, $\xi_0=(q_0,s_0)=(0,0)$, and $a_0=0$.
\FOR{$t=0,\dots,T-1$}
  \STATE Build observation $\mathbf{o}_t$ and update memory $\mathbf{m}_t$ from the current signals $(\rho^{(t)}, \Delta\rho^{(t)}, \{\tau_{t,w}, \gamma_{t,w}, \phi_{t,w}\}_{w\in\mathcal{U}_f}, \eta_t, \xi_t, a_t)$.
  \STATE Based on these signals, the Coordinator selects one round-level strategy and assigns each worker a role in $\{\textsc{PR},\textsc{CA},\textsc{SN},\textsc{IN}\}$.
  \STATE The Coordinator validates and aggregates accepted worker actions into one batch $\Delta\tilde{\mathbf{R}}^{(t+1)}$.
  \STATE Update $\tilde{\mathbf{R}}\leftarrow \tilde{\mathbf{R}} \cup \Delta\tilde{\mathbf{R}}^{(t+1)}$ and form the current poisoned matrix $[\mathbf{R};\,\tilde{\mathbf{R}}]$.
  \STATE Update worker signals, rank feedback, and environment signals $(\eta_{t+1}, \xi_{t+1}, a_{t+1})$.
  \STATE Update victim-family prediction, worker states, and Coordinator memory.
  \IF{the stopping condition is met}
    \STATE \textbf{break}
  \ENDIF
\ENDFOR
\STATE $\mathbf{R}'\leftarrow [\mathbf{R};\,\tilde{\mathbf{R}}]$.
\STATE \textbf{return} $\mathbf{R}'$
\end{algorithmic}
\end{minipage}%
}
\end{algorithm}

\subsection{Coordinator}
\label{sec:coordinator}
At round $t$, the Coordinator reads the current observation $\mathbf{o}_t$, updates memory $\mathbf{m}_t$, and computes the control signals directly from workers' actions and environment feedback. 

\sstitle{Worker signals}
These signals are maintained for each worker and summarize how safe that worker is.
\begin{itemize}

\item \textbf{Trust and risk scores.} AGAS maintains a trust score $\tau_{t,w}$ and a risk score $\gamma_{t,w}$ for each worker $w$. After an accepted action on item $i$ with rating $r_{t,w}$, it computes the deviation from the current item bias as: $d_{t,w}=|r_{t,w}-b_i|$. Trust increases only when the action remains close to the item bias:
\begin{equation}
\tau_{t+1,w}
=
\max\!\left(
0,\tau_{t,w}+\mathbb{I}[d_{t,w}\le 1]
\right).
\label{eq:trust_update}
\end{equation}
Risk increases when the action deviates from the bias, with an additional penalty for extreme actions from low-trust workers:
\begin{equation}
\begin{aligned}
\gamma_{t+1,w}
=\;&\max\!\left(
0,\gamma_{t,w}
+0.5\,\mathbb{I}[d_{t,w}>1]\right.\\
&\left.\qquad\quad
+0.5\,\mathbb{I}[d_{t,w}>1.5\land\tau_{t,w}<1]
\right).
\end{aligned}
\label{eq:risk_update}
\end{equation}
If worker $w$ is inactive in round $t$, risk decays as:
\begin{equation}
\gamma_{t+1,w}=\max\!\left(0,\gamma_{t,w}-0.5\right).
\label{eq:risk_decay}
\end{equation}

\item \textbf{Validator score.} $\phi_{t,w}$ measures whether worker $w$'s whole profile appears structurally suspicious:
\begin{equation}
\phi_{t,w}
=
\min\!\left(
1,\ 0.4\,e_{t,w}+0.3\,h_{t,w}+0.3\,o_{t,w}
\right),
\label{eq:profile_validator}
\end{equation}
where $e_{t,w}$ is the frequency of extreme ratings (near the two ends of the scale), $h_{t,w}$ indicates whether $w$ has given the target item a clear positive rating (near the positive end of the scale), and $o_{t,w}$ measures fake-user overlap, i.e., the fraction of $w$'s rated items also rated by at least two other fake users. Unlike $\tau_{t,w}$ and $\gamma_{t,w}$, which track recent actions, $\phi_{t,w}$ evaluates profile-level suspicion.
\end{itemize}

\sstitle{Environment signals}
These signals summarize how the victim model and platform are reacting to the campaign at each round. Unlike worker signals, which describe the status of individual fake users, environment signals capture global progress, possible defensive behavior, and campaign-level risk.

\begin{itemize}
\item \textbf{Rank feedback.} $\rho^{(t)}$ and $\Delta\rho^{(t)}$ denote the current mean target rank and its recent change. Together, they indicate whether the campaign is effectively promoting the target item or has started to stall. A consistent improvement in $\rho^{(t)}$ suggests that the current strategy is working, whereas a weak or negative $\Delta\rho^{(t)}$ may suggest that AGAS should adjust its strategy, increase exploration, or switch workers to safer roles.

\item \textbf{Acceptance rate.} $\eta_t$ is the recent ratio of accepted to attempted actions by active workers. It measures whether injected interactions are still being incorporated by the victim system. A sudden drop in $\eta_t$ may indicate silent filtering, discounting, or defensive suppression, even without an explicit alert.

\item \textbf{Suppression signal.}
$\xi_t=(q_t,s_t)$ denotes the round-level suppression signal, where $q_t\in[0,1]$ is the overall suspicion score and $s_t$ is the campaign-level suppression streak. The streak $s_t$ increases when the current round shows signs of filtering, discounting, or weak target-rank improvement. Let $\mathcal{W}_t$ be the active fake workers at round $t$. Let $\mathcal{A}^{\mathrm{try}}_t$, $\mathcal{A}^{\mathrm{drop}}_t$, $\mathcal{A}^{\mathrm{acc}}_t$, and $\mathcal{A}^{\mathrm{tar}}_t$ denote the attempted, dropped, accepted, and target-related actions in that round. We define:
\begin{equation}
\begin{aligned}
\hat d_t &= 
\frac{|\mathcal{A}^{\mathrm{drop}}_t|}
{\max(1,|\mathcal{A}^{\mathrm{try}}_t|)},\\
\hat\delta_t &=
\min\!\left(
1,\frac{1}{2|\mathcal{A}^{\mathrm{acc}}_t|}
\sum_{a\in\mathcal{A}^{\mathrm{acc}}_t}\delta(a)
\right),\\
\hat m_t &=
\frac{1}{\max(1,|\mathcal{A}^{\mathrm{tar}}_t|)}
\sum_{a\in\mathcal{A}^{\mathrm{tar}}_t}
\mathbb{I}[\Delta\rho^{\mathrm{tar}}(a)<\epsilon_\rho],\\
\hat s_t &= \min(1,s_t/3).
\end{aligned}
\label{eq:round_suppression_terms}
\end{equation}
Here, $\hat d_t$ is the dropped-action ratio, $\hat\delta_t$ is the normalized discount magnitude, $\hat m_t$ is the weak target-movement ratio, and $\hat s_t$ is the normalized suppression streak. $\delta(a)$ is the estimated discount applied to accepted action $a$, $\Delta\rho^{\mathrm{tar}}(a)$ is the target-rank improvement after action $a$, and $\epsilon_\rho$ is the minimum improvement threshold. We also compute a round-level group-overlap score:
\begin{equation}
\begin{aligned}
A_{t,w} &= \{\text{items rated by worker } w \text{ at round } t\},\\
g_t &= 
\max_{u,v\in\mathcal{W}_t,\ u\neq v}
\frac{|A_{t,u}\cap A_{t,v}|}{|A_{t,u}\cup A_{t,v}|}.
\end{aligned}
\label{eq:round_group_overlap}
\end{equation}
This captures whether fake workers are acting too similarly in the same round. To avoid over-tuning the suppression score, we give the five components equal weight and use a simple linear combination to compute $q_t$. The final round-level suspicion score is:
\begin{equation}
q_t
=
\min\!\Bigl(
1,\ 
0.2\hat d_t
+0.2\hat\delta_t
+0.2\hat m_t
+0.2\hat s_t
+0.2g_t
\Bigr).
\label{eq:round_suppression_score}
\end{equation}
A large $q_t$ means the campaign is likely being filtered, discounted, or detected at the environment level.

\item \textbf{Alert flag.}
$a_t\in\{0,1\}$ is the round-level alert flag. AGAS sets $a_t=1$ when the current round creates an abnormal target spike, unusually high worker overlap, or a sharp drop in action acceptance. This is a coarse practical trigger rather than a formal detector. It indicates a high-risk state and encourages the Coordinator to slow down, pause, or switch to safer strategies.
\end{itemize}

Before the campaign begins, AGAS evaluates the clean victim once to obtain the initial target rank $\rho^{(0)}$. Since no fake action has been injected, $\Delta\rho^{(0)}=0$. For each worker $w$, the \textit{Worker signals} are initialized as $\tau_{0,w}=0$, $\gamma_{0,w}=0$, and $\phi_{0,w}=0$. The round-level \textit{Environment signals} also start from neutral values: $\eta_0=1$, $\xi_0=(q_0,s_0)=(0,0)$, and $a_0=0$. The \textit{Worker signals} indicate which fake users are still effective or becoming risky. The \textit{Environment signals} summarize whether the victim platform appears to be accepting the campaign or entering a defensive state. Based on this feedback, the Coordinator decides whether to explore, attack, slow down, pause, or switch roles before selecting the next strategy.

\subsection{Workers}
\label{sec:workers}
Each worker keeps its own memory. This lets each worker decide what item to rate next from its own action history. Group-level coordination is still maintained by the Coordinator. The worker memory supports item-level decisions.

\providecommand{\bestval}[1]{\textbf{#1}}
\providecommand{\secondval}[1]{\underline{#1}}

\begin{table*}[t]
\centering
\tiny
\setlength{\tabcolsep}{1.6pt}
\renewcommand{\arraystretch}{1.15}
\caption{Unpopular-target promotion on embedding-based (top) and graph-based (bottom) victims. Mean$\pm$95\%~CI over 5 runs ($\times 10^{3}$). \textbf{Best} / \underline{second-best}.}
\label{tab:bench_unpop}
\resizebox{0.88\textwidth}{!}{%
\begin{tabular}{@{}l|cccc|cccc|cccc|cccc|cccc@{}}
\toprule
\multirow{3}{*}{Method} &
\multicolumn{4}{c|}{ML-100K} &
\multicolumn{4}{c|}{ML-1M} &
\multicolumn{4}{c|}{Amazon} &
\multicolumn{4}{c|}{Genome 2021} &
\multicolumn{4}{c}{Netflix} \\
\cmidrule(lr){2-5}\cmidrule(lr){6-9}\cmidrule(lr){10-13}\cmidrule(lr){14-17}\cmidrule(lr){18-21}
& \multicolumn{2}{c|}{MF (BPR)} & \multicolumn{2}{c|}{NeuMF}
& \multicolumn{2}{c|}{GMF} & \multicolumn{2}{c|}{NCF}
& \multicolumn{2}{c|}{MF (BPR)} & \multicolumn{2}{c|}{NCF}
& \multicolumn{2}{c|}{GMF} & \multicolumn{2}{c|}{NeuMF}
& \multicolumn{2}{c|}{MF (BPR)} & \multicolumn{2}{c}{NeuMF} \\
\cmidrule(lr){2-3}\cmidrule(lr){4-5}\cmidrule(lr){6-7}\cmidrule(lr){8-9}\cmidrule(lr){10-11}\cmidrule(lr){12-13}\cmidrule(lr){14-15}\cmidrule(lr){16-17}\cmidrule(lr){18-19}\cmidrule(lr){20-21}
& HR@10 & NDCG@10 & HR@10 & NDCG@10 & HR@10 & NDCG@10 & HR@10 & NDCG@10 & HR@10 & NDCG@10 & HR@10 & NDCG@10 & HR@10 & NDCG@10 & HR@10 & NDCG@10 & HR@10 & NDCG@10 & HR@10 & NDCG@10 \\
\midrule
NoneAttack
 & 1.8$\pm$0.2 & 0.7$\pm$0.5 & 2.2$\pm$0.2 & 0.9$\pm$0.6
 & 0.9$\pm$0.1 & 0.4$\pm$0.3 & 0.8$\pm$0.1 & 0.3$\pm$0.2
 & 0.4$\pm$0.1 & 0.1$\pm$0.2 & 0.5$\pm$0.1 & 0.2$\pm$0.2
 & 0.2$\pm$0.1 & 0.1$\pm$0.2 & 0.5$\pm$0.1 & 0.2$\pm$0.2
 & 0.6$\pm$0.1 & 0.2$\pm$0.2 & 0.7$\pm$0.1 & 0.3$\pm$0.2 \\
RandomAttack~\cite{omahony2005attack_types}
 & 4.1$\pm$0.3 & 1.6$\pm$0.6 & 4.7$\pm$0.3 & 1.8$\pm$0.6
 & 2.4$\pm$0.2 & 1.0$\pm$0.5 & 2.0$\pm$0.2 & 0.8$\pm$0.4
 & 1.2$\pm$0.1 & 0.5$\pm$0.3 & 1.3$\pm$0.1 & 0.5$\pm$0.3
 & 0.7$\pm$0.1 & 0.3$\pm$0.2 & 1.3$\pm$0.1 & 0.5$\pm$0.3
 & 1.8$\pm$0.2 & 0.7$\pm$0.4 & 2.0$\pm$0.2 & 0.8$\pm$0.5 \\
BandwagonAttack~\cite{omahony2005attack_types}
 & 6.2$\pm$0.3 & 2.5$\pm$0.7 & 6.8$\pm$0.3 & 2.7$\pm$0.8
 & 3.4$\pm$0.2 & 1.4$\pm$0.6 & 2.9$\pm$0.2 & 1.2$\pm$0.5
 & 2.0$\pm$0.1 & 0.8$\pm$0.4 & 2.2$\pm$0.1 & 0.9$\pm$0.4
 & 1.2$\pm$0.1 & 0.5$\pm$0.3 & 2.0$\pm$0.1 & 0.8$\pm$0.4
 & 2.7$\pm$0.2 & 1.1$\pm$0.5 & 3.0$\pm$0.2 & 1.2$\pm$0.5 \\
AUSH~\cite{lin2020aush}
 & 10.6$\pm$0.6 & 4.3$\pm$1.2 & 10.0$\pm$0.5 & 4.1$\pm$1.1
 & 6.5$\pm$0.4 & 2.6$\pm$0.9 & 4.3$\pm$0.3 & 1.8$\pm$0.7
 & 3.3$\pm$0.3 & 1.3$\pm$0.6 & 3.1$\pm$0.3 & 1.2$\pm$0.6
 & 1.9$\pm$0.2 & 0.8$\pm$0.5 & 2.8$\pm$0.3 & 1.1$\pm$0.6
 & 4.2$\pm$0.3 & 1.7$\pm$0.7 & 4.5$\pm$0.3 & 1.8$\pm$0.8 \\
PoisonRec~\cite{fang2020poisonrec}
 & 13.4$\pm$0.7 & 5.3$\pm$1.6 & 11.5$\pm$0.6 & 4.7$\pm$1.5
 & 7.6$\pm$0.5 & 3.0$\pm$1.1 & 5.5$\pm$0.4 & 2.1$\pm$0.9
 & \secondval{5.8$\pm$0.5} & \secondval{2.3$\pm$1.0} & 4.8$\pm$0.3 & 1.9$\pm$0.8
 & 2.3$\pm$0.3 & 0.9$\pm$0.6 & 3.2$\pm$0.3 & 1.3$\pm$0.7
 & 5.6$\pm$0.4 & 2.2$\pm$1.0 & 5.7$\pm$0.4 & 2.3$\pm$1.0 \\
PGA~\cite{li2016pga_cf_poisoning}
 & 11.5$\pm$0.6 & 4.6$\pm$1.3 & 12.2$\pm$0.6 & 5.0$\pm$1.4
 & 8.4$\pm$0.5 & 3.4$\pm$1.1 & 6.0$\pm$0.4 & 2.4$\pm$0.9
 & 4.3$\pm$0.3 & 1.7$\pm$0.8 & 4.0$\pm$0.3 & 1.6$\pm$0.7
 & 2.8$\pm$0.2 & 1.1$\pm$0.6 & 3.8$\pm$0.3 & 1.5$\pm$0.7
 & 5.5$\pm$0.4 & 2.2$\pm$1.0 & 6.0$\pm$0.5 & 2.4$\pm$1.0 \\
AgentSA~\cite{gu2026llm_agent_shilling_wsdm}
 & 12.7$\pm$0.6 & 5.2$\pm$1.4 & \secondval{12.5$\pm$0.6} & \secondval{5.1$\pm$1.4}
 & 8.2$\pm$0.4 & 3.3$\pm$1.0 & \secondval{7.2$\pm$0.4} & \secondval{2.9$\pm$1.0}
 & 5.6$\pm$0.3 & 2.2$\pm$0.9 & 4.9$\pm$0.3 & 1.9$\pm$0.8
 & \secondval{2.9$\pm$0.2} & \secondval{1.1$\pm$0.6} & 4.2$\pm$0.3 & 1.7$\pm$0.7
 & \secondval{6.3$\pm$0.4} & \secondval{2.5$\pm$1.0} & 6.5$\pm$0.4 & 2.6$\pm$1.0 \\
AgentAttack~\cite{li2026agentattack}
 & \secondval{13.6$\pm$0.7} & \secondval{5.5$\pm$1.5} & 12.1$\pm$0.6 & 4.9$\pm$1.4
 & \secondval{8.8$\pm$0.5} & \secondval{3.5$\pm$1.1} & 6.8$\pm$0.4 & 2.7$\pm$1.0
 & 5.7$\pm$0.3 & 2.2$\pm$0.9 & \secondval{5.2$\pm$0.3} & \secondval{2.0$\pm$0.9}
 & 2.7$\pm$0.2 & 1.0$\pm$0.6 & \secondval{4.5$\pm$0.3} & \secondval{1.8$\pm$0.8}
 & 6.1$\pm$0.4 & 2.4$\pm$1.0 & \secondval{6.9$\pm$0.5} & \secondval{2.8$\pm$1.0} \\
\rowcolor{gray!15}
\textbf{AGAS (Ours)}
 & \bestval{40.0$\pm$0.2} & \bestval{16.1$\pm$0.3} & \bestval{35.0$\pm$0.2} & \bestval{14.2$\pm$0.3}
 & \bestval{22.6$\pm$0.1} & \bestval{9.0$\pm$0.2} & \bestval{17.6$\pm$0.1} & \bestval{7.1$\pm$0.2}
 & \bestval{16.1$\pm$0.1} & \bestval{6.3$\pm$0.2} & \bestval{15.0$\pm$0.1} & \bestval{5.9$\pm$0.2}
 & \bestval{8.2$\pm$0.1} & \bestval{3.2$\pm$0.2} & \bestval{11.5$\pm$0.1} & \bestval{4.6$\pm$0.2}
 & \bestval{18.2$\pm$0.1} & \bestval{7.3$\pm$0.2} & \bestval{19.6$\pm$0.1} & \bestval{7.9$\pm$0.2} \\
\midrule
\textbf{Improvement}
 & \bestval{187.8\%} & \bestval{182.5\%} & \bestval{186.9\%} & \bestval{184.0\%}
 & \bestval{162.8\%} & \bestval{164.7\%} & \bestval{155.1\%} & \bestval{153.6\%}
 & \bestval{177.6\%} & \bestval{173.9\%} & \bestval{172.7\%} & \bestval{168.2\%}
 & \bestval{192.9\%} & \bestval{190.9\%} & \bestval{161.4\%} & \bestval{155.6\%}
 & \bestval{198.4\%} & \bestval{192.0\%} & \bestval{192.5\%} & \bestval{192.6\%} \\
\midrule
\multicolumn{21}{@{}l}{} \\[-0.6em]
& \multicolumn{4}{c|}{} &
\multicolumn{4}{c|}{} &
\multicolumn{4}{c|}{} &
\multicolumn{4}{c|}{} &
\multicolumn{4}{c}{Douban Movie} \\
\cmidrule(lr){2-5}\cmidrule(lr){6-9}\cmidrule(lr){10-13}\cmidrule(lr){14-17}\cmidrule(lr){18-21}
& \multicolumn{2}{c|}{NGCF} & \multicolumn{2}{c|}{LightGCN}
& \multicolumn{2}{c|}{SimGCL} & \multicolumn{2}{c|}{XSimGCL}
& \multicolumn{2}{c|}{EGCF} & \multicolumn{2}{c|}{LightCCF}
& \multicolumn{2}{c|}{NGCF} & \multicolumn{2}{c|}{LightGCN}
& \multicolumn{2}{c|}{NGCF} & \multicolumn{2}{c}{LightGCN} \\
\cmidrule(lr){2-3}\cmidrule(lr){4-5}\cmidrule(lr){6-7}\cmidrule(lr){8-9}\cmidrule(lr){10-11}\cmidrule(lr){12-13}\cmidrule(lr){14-15}\cmidrule(lr){16-17}\cmidrule(lr){18-19}\cmidrule(lr){20-21}
& HR@10 & NDCG@10 & HR@10 & NDCG@10 & HR@10 & NDCG@10 & HR@10 & NDCG@10 & HR@10 & NDCG@10 & HR@10 & NDCG@10 & HR@10 & NDCG@10 & HR@10 & NDCG@10 & HR@10 & NDCG@10 & HR@10 & NDCG@10 \\
\midrule
NoneAttack
 & 1.9$\pm$0.2 & 0.8$\pm$0.5 & 2.4$\pm$0.2 & 1.0$\pm$0.6
 & 0.8$\pm$0.1 & 0.3$\pm$0.2 & 0.9$\pm$0.1 & 0.4$\pm$0.3
 & 0.4$\pm$0.1 & 0.2$\pm$0.2 & 0.5$\pm$0.1 & 0.2$\pm$0.2
 & 0.2$\pm$0.1 & 0.1$\pm$0.2 & 0.3$\pm$0.1 & 0.1$\pm$0.2
 & 0.7$\pm$0.1 & 0.3$\pm$0.2 & 0.6$\pm$0.1 & 0.2$\pm$0.2 \\
RandomAttack~\cite{omahony2005attack_types}
 & 4.5$\pm$0.3 & 1.8$\pm$0.6 & 5.0$\pm$0.3 & 2.0$\pm$0.7
 & 1.8$\pm$0.2 & 0.7$\pm$0.4 & 2.2$\pm$0.2 & 0.9$\pm$0.5
 & 1.1$\pm$0.1 & 0.4$\pm$0.3 & 1.2$\pm$0.1 & 0.5$\pm$0.3
 & 0.6$\pm$0.1 & 0.2$\pm$0.2 & 0.6$\pm$0.1 & 0.2$\pm$0.2
 & 2.0$\pm$0.2 & 0.8$\pm$0.5 & 1.9$\pm$0.2 & 0.7$\pm$0.4 \\
BandwagonAttack~\cite{omahony2005attack_types}
 & 5.9$\pm$0.3 & 2.4$\pm$0.7 & 6.7$\pm$0.3 & 2.7$\pm$0.8
 & 2.4$\pm$0.2 & 1.0$\pm$0.5 & 2.9$\pm$0.2 & 1.2$\pm$0.6
 & 1.8$\pm$0.2 & 0.7$\pm$0.4 & 1.9$\pm$0.2 & 0.8$\pm$0.4
 & 1.1$\pm$0.1 & 0.4$\pm$0.3 & 1.0$\pm$0.1 & 0.4$\pm$0.3
 & 2.8$\pm$0.2 & 1.1$\pm$0.6 & 2.7$\pm$0.2 & 1.1$\pm$0.5 \\
GSPAttack~\cite{nguyen2023gspattack}
 & 11.0$\pm$0.6 & 4.5$\pm$1.3 & 11.5$\pm$0.6 & 4.7$\pm$1.3
 & 4.8$\pm$0.3 & 1.9$\pm$0.8 & 5.8$\pm$0.4 & 2.3$\pm$0.9
 & 3.2$\pm$0.3 & 1.3$\pm$0.6 & 3.4$\pm$0.3 & 1.4$\pm$0.6
 & 2.0$\pm$0.2 & 0.8$\pm$0.5 & 1.8$\pm$0.2 & 0.7$\pm$0.5
 & 5.0$\pm$0.3 & 2.0$\pm$0.8 & 4.8$\pm$0.3 & 1.9$\pm$0.8 \\
TargetedAttack~\cite{guo2023targeted_gnn_shilling}
 & 11.9$\pm$0.6 & 4.8$\pm$1.3 & 12.6$\pm$0.6 & 5.1$\pm$1.4
 & 5.4$\pm$0.4 & 2.2$\pm$0.8 & 6.4$\pm$0.4 & 2.5$\pm$1.0
 & 4.7$\pm$0.3 & 1.9$\pm$0.7 & 4.5$\pm$0.3 & 1.8$\pm$0.7
 & 2.5$\pm$0.2 & 1.0$\pm$0.6 & 2.2$\pm$0.2 & 0.9$\pm$0.5
 & 5.4$\pm$0.4 & 2.2$\pm$0.8 & 5.2$\pm$0.3 & 2.1$\pm$0.8 \\
CLeaR~\cite{wang2024unveiling_contrastive_poisoning}
 & 13.1$\pm$0.7 & 5.4$\pm$1.4 & 13.9$\pm$0.7 & 5.7$\pm$1.5
 & 6.4$\pm$0.4 & 2.6$\pm$0.9 & 7.7$\pm$0.5 & 3.1$\pm$1.0
 & 4.6$\pm$0.3 & 1.8$\pm$0.7 & 5.2$\pm$0.4 & 2.1$\pm$0.8
 & 2.9$\pm$0.3 & 1.2$\pm$0.6 & 2.8$\pm$0.2 & 1.1$\pm$0.6
 & 5.8$\pm$0.4 & 2.3$\pm$0.9 & 5.5$\pm$0.3 & 2.2$\pm$0.9 \\
AgentSA~\cite{gu2026llm_agent_shilling_wsdm}
 & 13.8$\pm$0.6 & 5.6$\pm$1.4 & \secondval{14.2$\pm$0.7} & \secondval{5.8$\pm$1.5}
 & 6.1$\pm$0.4 & 2.4$\pm$0.8 & \secondval{8.4$\pm$0.5} & \secondval{3.3$\pm$1.0}
 & 4.5$\pm$0.3 & 1.8$\pm$0.7 & \secondval{5.4$\pm$0.3} & \secondval{2.2$\pm$0.8}
 & 2.8$\pm$0.2 & 1.1$\pm$0.6 & \secondval{2.9$\pm$0.2} & \secondval{1.1$\pm$0.6}
 & 5.4$\pm$0.3 & 2.2$\pm$0.8 & \secondval{5.8$\pm$0.4} & \secondval{2.3$\pm$0.9} \\
AgentAttack~\cite{li2026agentattack}
 & \secondval{14.4$\pm$0.7} & \secondval{5.9$\pm$1.5} & 14.0$\pm$0.7 & 5.6$\pm$1.5
 & \secondval{6.6$\pm$0.4} & \secondval{2.6$\pm$0.9} & 8.0$\pm$0.5 & 3.1$\pm$1.0
 & \secondval{4.9$\pm$0.3} & \secondval{1.9$\pm$0.7} & 5.0$\pm$0.3 & 2.0$\pm$0.8
 & \secondval{3.1$\pm$0.3} & \secondval{1.2$\pm$0.6} & 2.5$\pm$0.2 & 1.0$\pm$0.5
 & \secondval{6.0$\pm$0.4} & \secondval{2.4$\pm$0.9} & 5.5$\pm$0.3 & 2.2$\pm$0.9 \\
\rowcolor{gray!15}
\textbf{AGAS (Ours)}
 & \bestval{38.8$\pm$0.3} & \bestval{17.0$\pm$0.3} & \bestval{40.5$\pm$0.2} & \bestval{16.5$\pm$0.3}
 & \bestval{16.8$\pm$0.1} & \bestval{6.7$\pm$0.2} & \bestval{20.7$\pm$0.1} & \bestval{8.2$\pm$0.2}
 & \bestval{13.1$\pm$0.1} & \bestval{5.2$\pm$0.2} & \bestval{13.8$\pm$0.1} & \bestval{5.4$\pm$0.2}
 & \bestval{8.9$\pm$0.1} & \bestval{3.4$\pm$0.2} & \bestval{7.8$\pm$0.1} & \bestval{3.1$\pm$0.2}
 & \bestval{17.0$\pm$0.1} & \bestval{6.8$\pm$0.2} & \bestval{16.9$\pm$0.1} & \bestval{6.7$\pm$0.2} \\
\midrule
\textbf{Improvement}
 & \bestval{181.7\%} & \bestval{193.1\%} & \bestval{189.3\%} & \bestval{194.6\%}
 & \bestval{162.5\%} & \bestval{157.7\%} & \bestval{155.6\%} & \bestval{156.2\%}
 & \bestval{178.7\%} & \bestval{173.7\%} & \bestval{165.4\%} & \bestval{157.1\%}
 & \bestval{196.7\%} & \bestval{183.3\%} & \bestval{178.6\%} & \bestval{181.8\%}
 & \bestval{193.1\%} & \bestval{195.7\%} & \bestval{196.5\%} & \bestval{191.3\%} \\
\bottomrule
\end{tabular}%
}
\end{table*}

\sstitle{Profiler (PR)}
PR is the exploration role. It uses a few safe interactions on \emph{filler items}, i.e., popular or related non-target items that normal users would reasonably rate, to test whether the platform is accepting actions and whether suppression or throttling is emerging. When feedback suggests a graph-style victim, it prioritizes \emph{bridge items}, i.e., an intermediate item that links the target to real users through shared preferences, to find receptive users and expand the attack path.

\sstitle{Camouflageur (CA)}
CA is the stealth role. It passively rates filler items instead of aggressively pushing the target, making the fake-user history look more like normal activity than coordinated promotion. This role dilutes suspicious behavior, helps rebuild trust after risky rounds, and keeps the fake-user pool active when the Coordinator slows down.

\sstitle{Sniper (SN)}
SN is the payload role, assigned to high-trust and low-risk workers because it creates the strongest rank movement but also the highest detection pressure. For embedding-based victims such as GMF, SN directly promotes the target item. For graph-based victims such as LightGCN, SN uses \emph{bridge items} discovered by PR to form useful two-hop paths between the target and receptive real-user neighborhoods, strengthening collaborative propagation through the interaction graph.

\sstitle{Inactive (IN)} The Inactive role means that the worker does not act in the current round. This role is necessary because a suspicious worker should disappear completely instead of continuing weak benign behavior.

\subsection{Strategy}
\label{sec:strategy}
For each round, the Coordinator selects one strategy and assigns one role to each worker.

\sstitle{1. Victim Probe} In the first few rounds, the Coordinator assigns some workers to PR to probe the victim family. If direct target-related actions improve the mean target rank $\rho^{(t)}$ or produce a favorable rank change $\Delta\rho^{(t)}$, the victim is treated as more likely embedding-based. Otherwise, it is treated as more likely graph-based. This is sufficient to guide later role allocation, as supported in \autoref{sec:ablation}.

\sstitle{2. Bridge Building} For graph-based victims, PR workers build a bridge-item pool by ranking non-target items according to how often they are rated by real users connected to the target. Items that produce favorable rank feedback for the target are added to the pool. These bridge items help SN create stronger two-hop paths to the target.

\sstitle{3. Warm-up} After probing, the Coordinator avoids early over-attack. It usually assigns one PR and a small number of CA so the campaign can build worker trust $\tau_{t,w}$ and collect clean feedback before using SN.

\sstitle{4. First Push} After warm-up, the Coordinator launches the first coordinated payload. A small number of SN workers attack together while at least one CA remains active as cover. This step aims to improve $\rho^{(t)}$ and $\Delta\rho^{(t)}$ without making the whole worker pool look synchronized.

\sstitle{5. Silent Slowdown} If the acceptance rate $\eta_t$ falls below the early baseline while no alert is raised $(a_t=0)$, the platform may be silently suppressing the attack. The Coordinator reduces campaign-level pressure by shifting only the cleanest high-trust workers with larger $\tau_{t,w}$ toward CA behavior, while others may keep baseline roles, including SN. This keeps AGAS less visible while it observes whether stronger attack is needed.

\sstitle{6. Profile Cleanup} If worker risk $\phi_{t,w}$ rises, suppression $\xi_t=(q_t,s_t)$ increases, or an alert appears $(a_t=1)$, the Coordinator sanitizes the active pool. It freezes the most suspicious SN, avoids profiles with strong fake-user overlap, and only nudges the cleanest IN into CA mode to preserve benign throughput. This strategy mainly removes detectable profiles.

\sstitle{7. Safe Replacement} If the environment enters a high-risk state, indicated by $a_t=1$ or a large round-level suspicion score $q_t$, the Coordinator substitutes suspicious workers with rested ones that have lower risk $\gamma_{t,w}$ and assigns these replacements to safer roles. This keeps the campaign operational while avoiding repeated use of the same exposed workers. The number of SN workers remains the same as before, but they are more likely to be trusted and less likely to be detected.

\sstitle{8. Main Attack} If no strong suspicious signals appear and the target is still far from top-K according to $\Delta\rho^{(t)}$, the best available worker is assigned the SN role. The Coordinator prefers workers with higher trust $\tau_{t,w}$ and lower risk $\gamma_{t,w}$, and continues pushing until the target reaches top-K or the environment signals indicate stronger defense.

\subsection{Complexity Analysis}
\label{sec:complexity}
Let $T$ denote the number of attack rounds, $|\mathcal{U}_f|$ the fake-user pool size, $L$ the per-user action budget, $V$ the victim-query cost, and $c_C,c_W$ the costs of one Coordinator and workers' calls. For embedding-based victims, $V=O(|\mathcal{U}_f|d)$, with $d$ the embedding dim. For graph-based victims, $V=O((|R|+|\tilde{R}^{(\le t)}|)dK)$ with $K$ propagation layers. The complexity of \autoref{alg:agas_end_to_end} is: $O\bigl(T(c_C+|\mathcal{U}_f|(c_W+L)+V)\bigr)$, with only $O(T)$ heavy Coordinator calls and $O(T|\mathcal{U}_f|)$ short worker calls. AGAS keeps the heavy LLM cost linear in $T$, whereas AgentAttack~\cite{li2026agentattack} first generates $C$ candidate profiles per worker and runs over a surrogate recommender, adding $O(|\mathcal{U}_f|C)$ generation plus surrogate retraining. The linear-in-$|\mathcal{U}_f|$ gap explains the token and runtime savings of AGAS in \autoref{sec:efficiency}.
\section{Experiments}
\label{sec:experiments}

\providecommand{\bestval}[1]{\textbf{#1}}
\providecommand{\secondval}[1]{\underline{#1}}

\begin{table*}[t]
\centering
\scriptsize
\setlength{\tabcolsep}{2.2pt}
\renewcommand{\arraystretch}{1.10}
\caption{Detection performance (Accuracy / Recall / Precision / F1). Lower values are harder to detect. Results are shown for PoisonRec, AgentSA, AgentAttack, and AGAS. \textbf{Best} / \underline{second} within each block.}
\label{tab:detect_mf}
\resizebox{0.8\textwidth}{!}{%
\begin{tabular}{@{}ll|cccc|cccc|cccc|cccc|cccc@{}}
\toprule
\multirow{2}{*}{Detector} & \multirow{2}{*}{Method} &
\multicolumn{4}{c|}{ML-100K} &
\multicolumn{4}{c|}{ML-1M} &
\multicolumn{4}{c|}{Netflix} &
\multicolumn{4}{c|}{Amazon} &
\multicolumn{4}{c}{Genome 2021} \\
\cmidrule(lr){3-6}\cmidrule(lr){7-10}\cmidrule(lr){11-14}\cmidrule(lr){15-18}\cmidrule(lr){19-22}
& & Accuracy & Recall & Precision & F1
  & Accuracy & Recall & Precision & F1
  & Accuracy & Recall & Precision & F1
  & Accuracy & Recall & Precision & F1
  & Accuracy & Recall & Precision & F1 \\
\midrule
\multirow{4}{*}{BaseDetect~\cite{williams2007defending}}
& PoisonRec~\cite{fang2020poisonrec}
& 0.942 & 0.920 & 0.915 & 0.917 & 0.938 & 0.915 & 0.910 & 0.912 & 0.932 & 0.910 & 0.905 & 0.907 & 0.926 & 0.905 & 0.900 & 0.902 & 0.930 & 0.907 & 0.903 & 0.905 \\
& AgentSA~\cite{gu2026llm_agent_shilling_wsdm}
& 0.932 & 0.905 & 0.895 & 0.900 & 0.928 & 0.900 & 0.890 & 0.895 & 0.920 & 0.892 & 0.882 & 0.887 & 0.914 & 0.885 & 0.875 & 0.880 & 0.918 & 0.888 & 0.878 & 0.883 \\
& AgentAttack~\cite{li2026agentattack}
& \secondval{0.918} & \secondval{0.892} & \secondval{0.880} & \secondval{0.886} & \secondval{0.913} & \secondval{0.886} & \secondval{0.875} & \secondval{0.880} & \secondval{0.905} & \secondval{0.878} & \secondval{0.866} & \secondval{0.872} & \secondval{0.900} & \secondval{0.872} & \secondval{0.860} & \secondval{0.866} & \secondval{0.904} & \secondval{0.875} & \secondval{0.864} & \secondval{0.870} \\
\rowcolor{gray!15}\cellcolor{white} & \textbf{AGAS (Ours)}
& \bestval{0.385} & \bestval{0.345} & \bestval{0.360} & \bestval{0.352} & \bestval{0.372} & \bestval{0.330} & \bestval{0.348} & \bestval{0.339} & \bestval{0.360} & \bestval{0.315} & \bestval{0.335} & \bestval{0.325} & \bestval{0.352} & \bestval{0.305} & \bestval{0.325} & \bestval{0.315} & \bestval{0.365} & \bestval{0.320} & \bestval{0.340} & \bestval{0.330} \\
\midrule
\multirow{4}{*}{DHAGCN~\cite{hao2023detection}}
& PoisonRec~\cite{fang2020poisonrec}
& 0.934 & 0.910 & 0.903 & 0.906 & 0.930 & 0.905 & 0.898 & 0.901 & 0.924 & 0.900 & 0.892 & 0.896 & 0.918 & 0.895 & 0.888 & 0.891 & \secondval{0.908} & \secondval{0.872} & \secondval{0.860} & \secondval{0.866} \\
& AgentSA~\cite{gu2026llm_agent_shilling_wsdm}
& 0.924 & 0.895 & 0.882 & 0.888 & 0.920 & 0.890 & 0.878 & 0.884 & 0.912 & 0.882 & 0.870 & 0.876 & 0.906 & 0.875 & 0.862 & 0.868 & 0.918 & 0.885 & 0.872 & 0.878 \\
& AgentAttack~\cite{li2026agentattack}
& \secondval{0.910} & \secondval{0.880} & \secondval{0.868} & \secondval{0.874} & \secondval{0.906} & \secondval{0.876} & \secondval{0.864} & \secondval{0.870} & \secondval{0.898} & \secondval{0.868} & \secondval{0.855} & \secondval{0.861} & \secondval{0.892} & \secondval{0.860} & \secondval{0.847} & \secondval{0.853} & 0.912 & 0.880 & 0.866 & 0.873 \\
\rowcolor{gray!15}\cellcolor{white} & \textbf{AGAS (Ours)}
& \bestval{0.372} & \bestval{0.330} & \bestval{0.350} & \bestval{0.340} & \bestval{0.360} & \bestval{0.315} & \bestval{0.338} & \bestval{0.326} & \bestval{0.348} & \bestval{0.300} & \bestval{0.325} & \bestval{0.312} & \bestval{0.340} & \bestval{0.290} & \bestval{0.315} & \bestval{0.302} & \bestval{0.352} & \bestval{0.305} & \bestval{0.330} & \bestval{0.317} \\
\midrule
\multirow{4}{*}{PCASelectUsers~\cite{mehta2009unsupervised}}
& PoisonRec~\cite{fang2020poisonrec}
& 0.922 & 0.790 & 0.945 & 0.861 & \secondval{0.892} & \secondval{0.748} & \secondval{0.920} & \secondval{0.825} & 0.909 & 0.770 & 0.935 & 0.845 & \secondval{0.880} & \secondval{0.730} & \secondval{0.912} & \secondval{0.811} & 0.906 & 0.765 & 0.932 & 0.840 \\
& AgentSA~\cite{gu2026llm_agent_shilling_wsdm}
& 0.912 & 0.775 & 0.935 & 0.848 & 0.906 & 0.765 & 0.930 & 0.839 & 0.898 & 0.755 & 0.925 & 0.831 & 0.892 & 0.745 & 0.920 & 0.823 & 0.896 & 0.750 & 0.922 & 0.827 \\
& AgentAttack~\cite{li2026agentattack}
& \secondval{0.898} & \secondval{0.760} & \secondval{0.920} & \secondval{0.832} & 0.910 & 0.770 & 0.935 & 0.845 & \secondval{0.884} & \secondval{0.738} & \secondval{0.910} & \secondval{0.815} & 0.896 & 0.750 & 0.922 & 0.828 & \secondval{0.882} & \secondval{0.732} & \secondval{0.908} & \secondval{0.812} \\
\rowcolor{gray!15}\cellcolor{white} & \textbf{AGAS (Ours)}
& \bestval{0.355} & \bestval{0.295} & \bestval{0.375} & \bestval{0.330} & \bestval{0.345} & \bestval{0.285} & \bestval{0.365} & \bestval{0.320} & \bestval{0.335} & \bestval{0.275} & \bestval{0.352} & \bestval{0.309} & \bestval{0.328} & \bestval{0.265} & \bestval{0.342} & \bestval{0.299} & \bestval{0.332} & \bestval{0.272} & \bestval{0.355} & \bestval{0.308} \\
\midrule
\multirow{4}{*}{GAGE~\cite{zhang2020gage}}
& PoisonRec~\cite{fang2020poisonrec}
& \secondval{0.805} & \secondval{0.630} & \secondval{0.840} & \secondval{0.720} & \secondval{0.795} & \secondval{0.615} & \secondval{0.830} & \secondval{0.707} & \secondval{0.780} & \secondval{0.600} & \secondval{0.818} & \secondval{0.692} & \secondval{0.770} & \secondval{0.585} & \secondval{0.805} & \secondval{0.678} & \secondval{0.782} & \secondval{0.598} & \secondval{0.815} & \secondval{0.690} \\
& AgentSA~\cite{gu2026llm_agent_shilling_wsdm}
& 0.930 & 0.905 & 0.940 & 0.922 & 0.925 & 0.900 & 0.935 & 0.917 & 0.918 & 0.895 & 0.930 & 0.912 & 0.912 & 0.890 & 0.925 & 0.907 & 0.920 & 0.895 & 0.930 & 0.912 \\
& AgentAttack~\cite{li2026agentattack}
& 0.880 & 0.755 & 0.892 & 0.818 & 0.870 & 0.738 & 0.882 & 0.804 & 0.860 & 0.722 & 0.870 & 0.788 & 0.852 & 0.708 & 0.862 & 0.778 & 0.864 & 0.722 & 0.872 & 0.792 \\
\rowcolor{gray!15}\cellcolor{white} & \textbf{AGAS (Ours)}
& \bestval{0.400} & \bestval{0.310} & \bestval{0.360} & \bestval{0.333} & \bestval{0.390} & \bestval{0.295} & \bestval{0.348} & \bestval{0.319} & \bestval{0.380} & \bestval{0.280} & \bestval{0.335} & \bestval{0.305} & \bestval{0.372} & \bestval{0.270} & \bestval{0.325} & \bestval{0.295} & \bestval{0.382} & \bestval{0.285} & \bestval{0.340} & \bestval{0.310} \\
\midrule
\multirow{4}{*}{MD-CBA~\cite{xu2023group_shilling}}
& PoisonRec~\cite{fang2020poisonrec}
& \secondval{0.790} & \secondval{0.615} & \secondval{0.830} & \secondval{0.707} & \secondval{0.780} & \secondval{0.600} & \secondval{0.820} & \secondval{0.693} & \secondval{0.765} & \secondval{0.585} & \secondval{0.808} & \secondval{0.679} & 0.870 & 0.738 & 0.875 & 0.802 & \secondval{0.768} & \secondval{0.585} & \secondval{0.805} & \secondval{0.678} \\
& AgentSA~\cite{gu2026llm_agent_shilling_wsdm}
& 0.935 & 0.910 & 0.945 & 0.927 & 0.930 & 0.905 & 0.940 & 0.922 & 0.923 & 0.900 & 0.935 & 0.917 & 0.917 & 0.895 & 0.930 & 0.912 & 0.925 & 0.900 & 0.935 & 0.917 \\
& AgentAttack~\cite{li2026agentattack}
& 0.860 & 0.738 & 0.875 & 0.802 & 0.852 & 0.722 & 0.866 & 0.788 & 0.842 & 0.708 & 0.858 & 0.776 & \secondval{0.755} & \secondval{0.572} & \secondval{0.798} & \secondval{0.668} & 0.846 & 0.710 & 0.860 & 0.779 \\
\rowcolor{gray!15}\cellcolor{white} & \textbf{AGAS (Ours)}
& \bestval{0.392} & \bestval{0.300} & \bestval{0.355} & \bestval{0.325} & \bestval{0.382} & \bestval{0.285} & \bestval{0.342} & \bestval{0.311} & \bestval{0.372} & \bestval{0.270} & \bestval{0.330} & \bestval{0.297} & \bestval{0.365} & \bestval{0.260} & \bestval{0.320} & \bestval{0.287} & \bestval{0.370} & \bestval{0.275} & \bestval{0.332} & \bestval{0.301} \\
\bottomrule
\end{tabular}%
}
\vspace{-1em}
\end{table*}

\begin{figure*}[t]
    \centering
    \begin{minipage}{0.49\textwidth}
        \centering
        \includegraphics[width=0.9\linewidth]{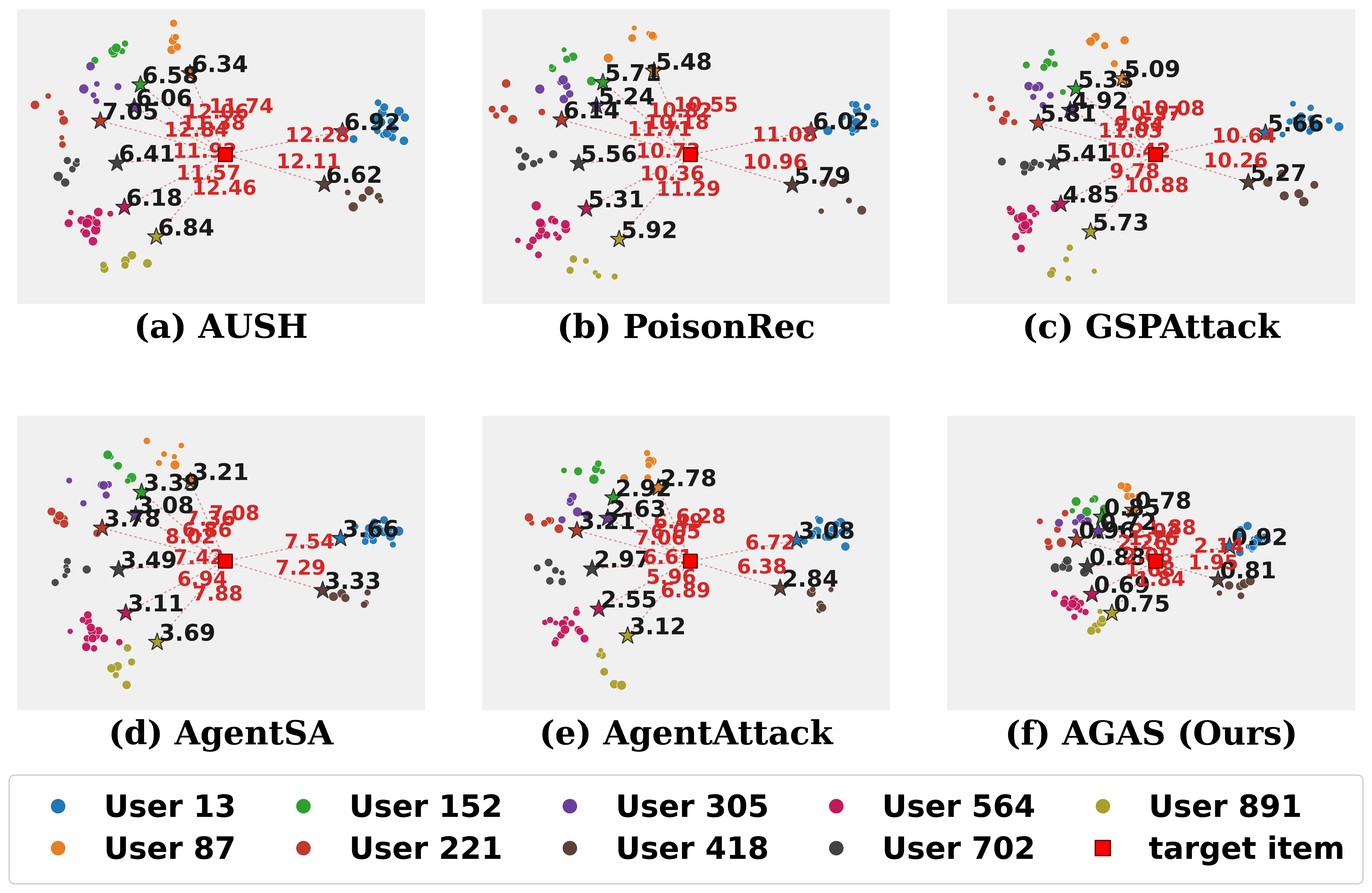}
    \end{minipage}\hfill
    \begin{minipage}{0.49\textwidth}
        \centering
        \includegraphics[width=0.9\linewidth]{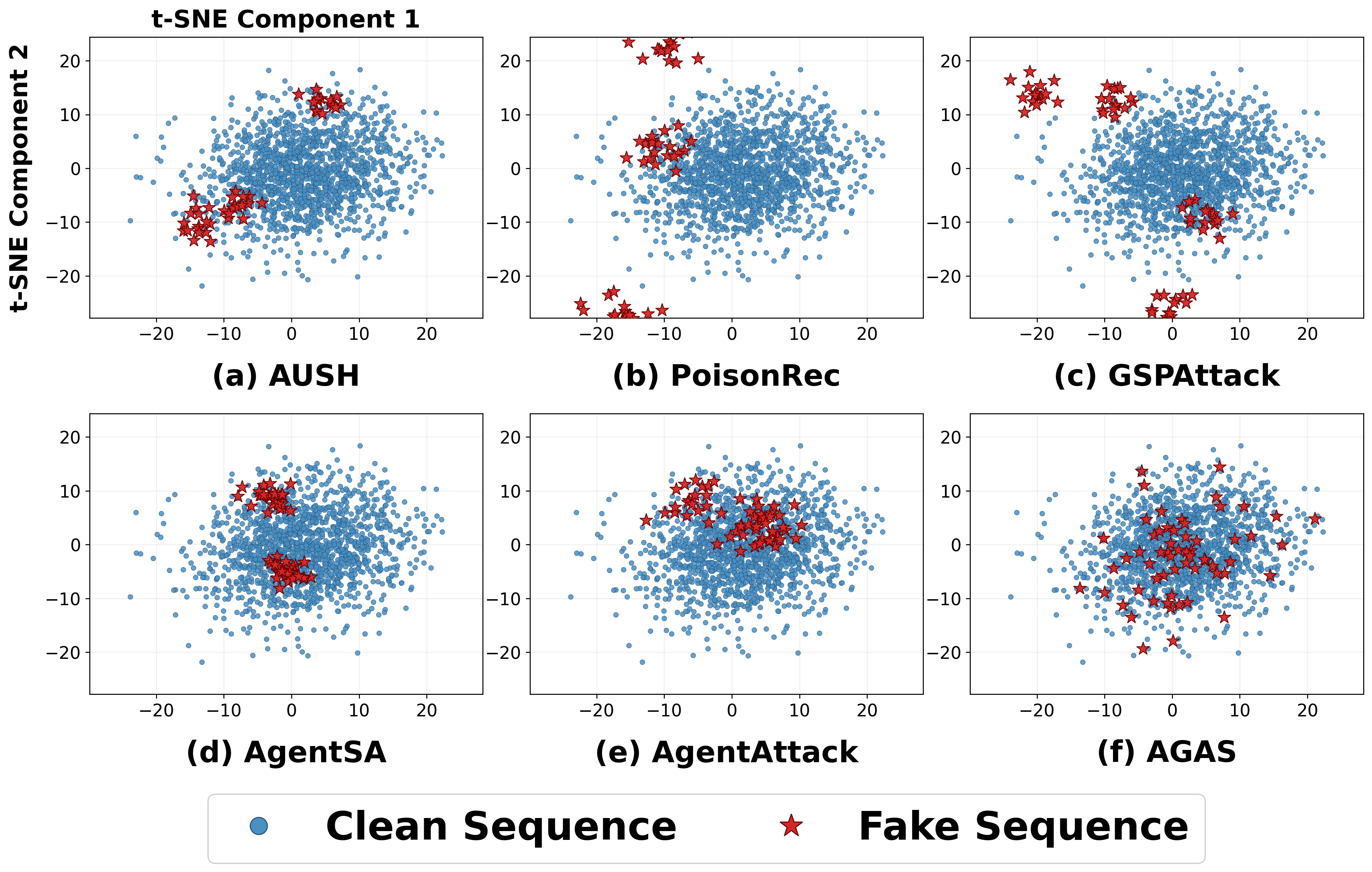}
    \end{minipage}
    \caption{t-SNE on ML-100K. \emph{Left:} red = distance to target, black = distance to ground-truth items (smaller is stealthier). \emph{Right:} clean (blue) vs.\ fake (star) user aggregate histories on poisoned victims. More scattered stars indicate better stealth.}
    \vspace{-.5em}
    \label{fig:stealth_tsne_combined}
\end{figure*}

\begin{figure}[t]
    \centering
    \includegraphics[width=0.85\linewidth]{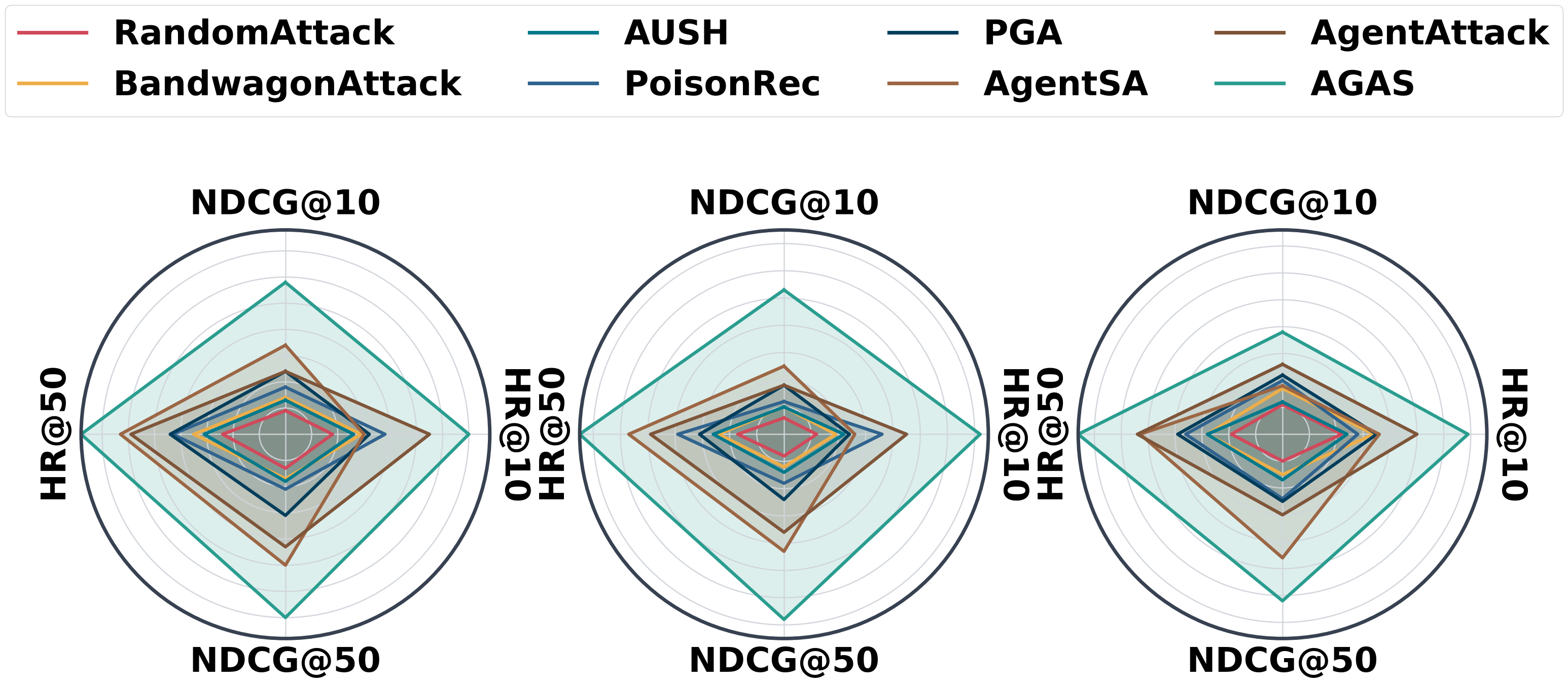}
    \caption{Summary of Head/Mid attack performance across representative victims: GMF (MF), LightGCN (graph convolution), and XSimGCL (graph contrastive).}
    \vspace{-.5em}
    \label{fig:popularity_regime}
\end{figure}

\begin{figure}[t]
    \centering
    \includegraphics[width=0.85\linewidth]{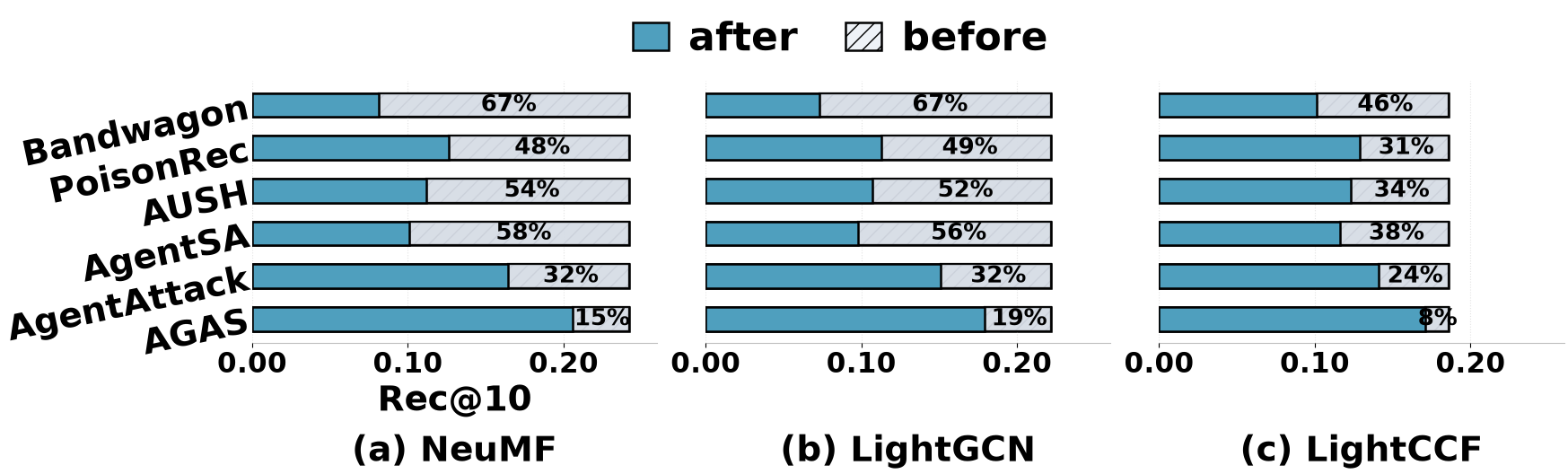}
    \caption{Benign Rec@10 on three representative victims. The gray segment shows the gap to NoneAttack. Higher is better.}
    \vspace{-.5em}
    \label{fig:detect_rec10_victim}
\end{figure}

\begin{figure}[t]
    \centering
    \includegraphics[width=0.85\linewidth]{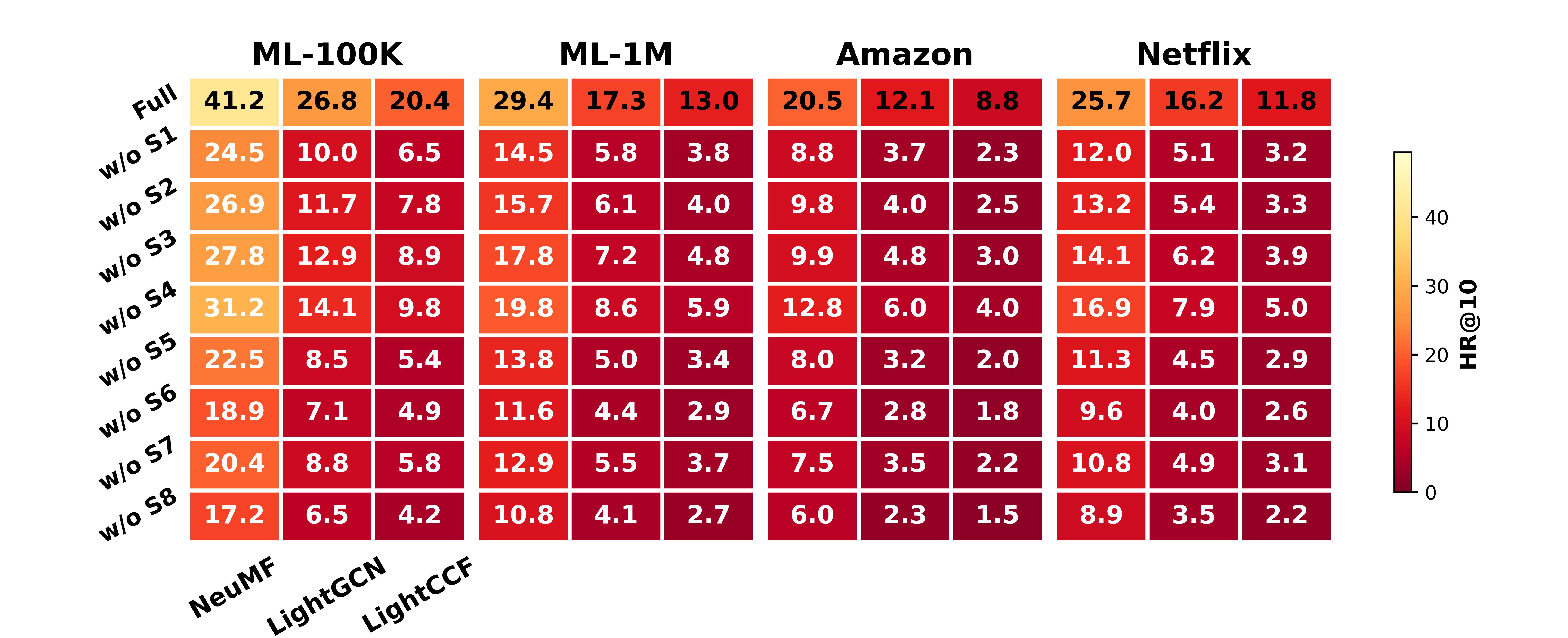}
    \caption{Strategy ablation heatmap on four datasets. Columns correspond to three representative victims from MF, graph convolution, and graph contrastive families.}
    \vspace{-1em}
    \label{fig:ablation_strategy_heatmap}
\end{figure}

\begin{figure}[t]
    \centering
    \includegraphics[width=0.85\linewidth]{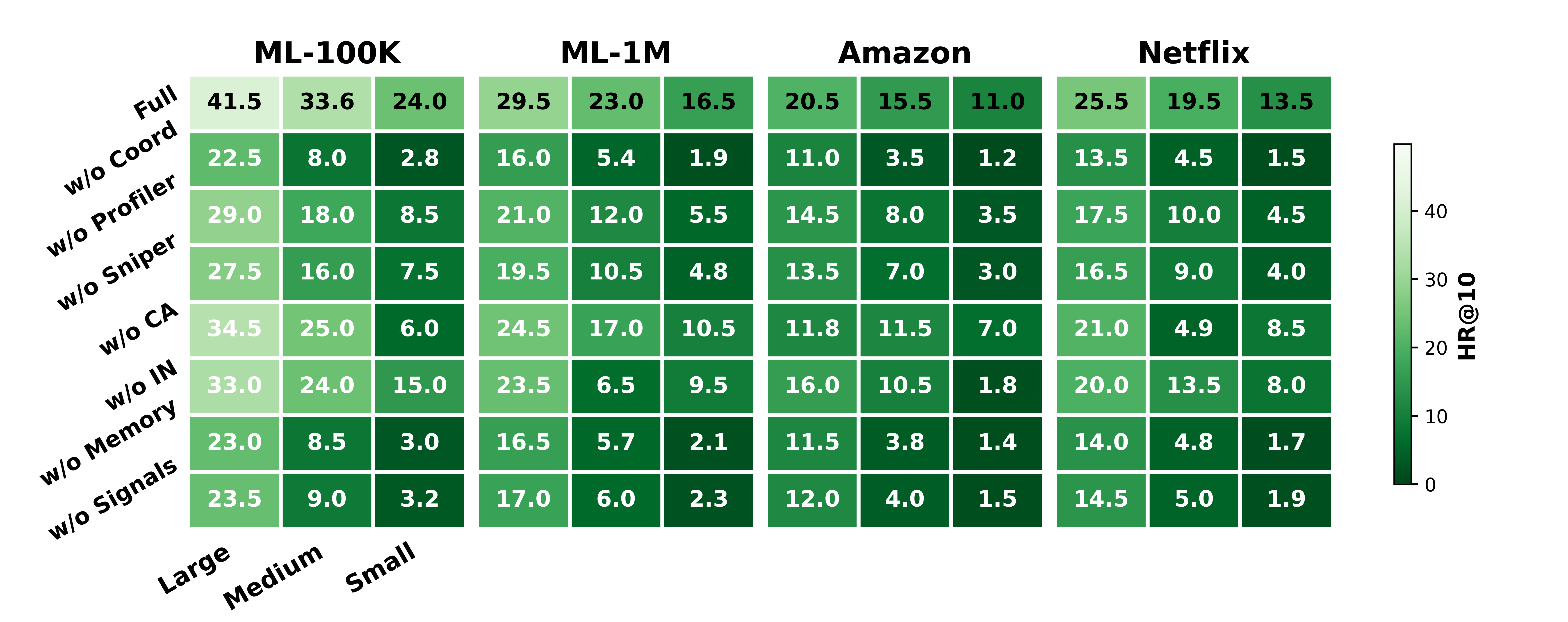}
    \caption{Component ablation across backbone size classes on four datasets. \emph{Large}, \emph{Medium}, and \emph{Small} denote the size of the underlying language model.}
    \vspace{-.5em}
    \label{fig:ablation_size_heatmap}
\end{figure}

\providecommand{\bestval}[1]{\textbf{#1}}
\providecommand{\secondval}[1]{\underline{#1}}

\begin{table}[t]
\centering
\scriptsize
\setlength{\tabcolsep}{3.6pt}
\renewcommand{\arraystretch}{1.10}
\caption{Backbone ablation grouped into size tiers, following the taxonomy of \cite{zhao2023llm_survey,wang2025slm_survey}. Values are mean$\pm$95\%~CI over 5 runs ($\times 10^{3}$). \textbf{Best} / \underline{second} within each model block.}
\label{tab:ablation_backbones}
\resizebox{0.6\linewidth}{!}{%
\begin{tabular}{@{}llccc@{}}
\toprule
Model & Method & HR@10 & NDCG@10 & Rec@50 \\
\midrule
\multicolumn{5}{@{}l}{\emph{Frontier (closed-source)}} \\
\midrule
\multirow{3}{*}{Claude Opus 4.5}
& AgentSA     & 34.4$\pm$1.6             & 14.5$\pm$1.4             & 82.8$\pm$1.9 \\
& AgentAttack & \secondval{35.2$\pm$1.6} & \secondval{14.9$\pm$1.4} & \secondval{84.3$\pm$1.9} \\
\rowcolor{gray!15}\cellcolor{white} & \textbf{AGAS}        & \bestval{41.5$\pm$0.5}   & \bestval{17.6$\pm$0.3}   & \bestval{95.0$\pm$1.1} \\
\midrule
\multirow{3}{*}{GPT-5.1}
& AgentSA     & \secondval{34.5$\pm$1.5} & \secondval{14.6$\pm$1.3} & \secondval{82.6$\pm$1.9} \\
& AgentAttack & 33.8$\pm$1.6             & 14.3$\pm$1.4             & 81.4$\pm$2.0 \\
\rowcolor{gray!15}\cellcolor{white} & \textbf{AGAS}        & \bestval{40.6$\pm$0.5}   & \bestval{17.2$\pm$0.3}   & \bestval{92.8$\pm$1.1} \\
\midrule
\multicolumn{5}{@{}l}{\emph{Large ($\geq$ 50B)}} \\
\midrule
\multirow{3}{*}{Llama 3.3 70B}
& AgentSA     & 30.6$\pm$1.7             & 12.9$\pm$1.5             & 74.6$\pm$2.0 \\
& AgentAttack & \secondval{31.2$\pm$1.7} & \secondval{13.2$\pm$1.5} & \secondval{75.9$\pm$2.0} \\
\rowcolor{gray!15}\cellcolor{white} & \textbf{AGAS}        & \bestval{37.0$\pm$0.6}   & \bestval{15.7$\pm$0.4}   & \bestval{84.8$\pm$1.3} \\
\midrule
\multirow{3}{*}{DeepSeek-V3}
& AgentSA     & \secondval{33.4$\pm$1.5} & \secondval{14.0$\pm$1.3} & \secondval{80.2$\pm$1.9} \\
& AgentAttack & 33.0$\pm$1.6             & 13.9$\pm$1.4             & 79.4$\pm$2.0 \\
\rowcolor{gray!15}\cellcolor{white} & \textbf{AGAS}        & \bestval{39.5$\pm$0.5}   & \bestval{16.7$\pm$0.3}   & \bestval{90.5$\pm$1.1} \\
\midrule
\multicolumn{5}{@{}l}{\emph{Medium ($\sim$ 10--30B)}} \\
\midrule
\multirow{3}{*}{Phi-4 (14B)}
& AgentSA     & 26.5$\pm$1.8             & \secondval{11.4$\pm$1.5} & \secondval{62.4$\pm$2.0} \\
& AgentAttack & \secondval{27.0$\pm$1.8} & 11.1$\pm$1.5             & 61.2$\pm$2.0 \\
\rowcolor{gray!15}\cellcolor{white} & \textbf{AGAS}        & \bestval{31.6$\pm$0.7}   & \bestval{13.3$\pm$0.5}   & \bestval{71.8$\pm$1.6} \\
\midrule
\multicolumn{5}{@{}l}{\emph{Small ($<$ 10B)}} \\
\midrule
\multirow{3}{*}{Gemma 2 2B}
& AgentSA     & 17.8$\pm$2.0             & 7.5$\pm$1.7              & 41.2$\pm$2.2 \\
& AgentAttack & \secondval{18.6$\pm$2.0} & \secondval{7.9$\pm$1.7}  & \secondval{42.8$\pm$2.2} \\
\rowcolor{gray!15}\cellcolor{white} & \textbf{AGAS}        & \bestval{22.4$\pm$1.0}   & \bestval{9.4$\pm$0.6}    & \bestval{51.0$\pm$2.0} \\
\bottomrule
\end{tabular}%
}
\vspace{-1em}
\end{table}


We design experiments to answer the following research questions (RQs):
\begin{compactitem}
\item \textbf{RQ1 (Performance).} How does AGAS perform across different victim models and popularity regimes?
\item \textbf{RQ2 (Stealthiness).} Can AGAS promote target items while preserving realistic user behavior?
\item \textbf{RQ3 (Detector).} How do anomaly detectors perform against AGAS?
\item \textbf{RQ4 (Ablation).} How does each component in AGAS contribute to overall performance?
\item \textbf{RQ5 (Efficiency).} How effectively and efficiently does AGAS scale as a benchmarking and stress-testing tool for RecSys?
\end{compactitem}

\subsection{Experimental Setup}

AGAS is a layered LLM-agent system inspired by ReAct~\cite{yao2023react} and Reflexion~\cite{shinn2023reflexion}. Fake-user workers use a ReAct-style loop to reason over profile states and execute rating actions, while the Coordinator uses a Reflexion-style loop to aggregate signals, update campaign memory, and adapt future roles and strategies.

\sstitle{Datasets}
We evaluate AGAS on public recommendation datasets that are commonly used for CF research and shilling-attack evaluation. These datasets cover a range of domains, sizes, and interaction patterns:
\begin{compactitem}
\item MovieLens-100K (ML-100K)~\cite{harper2015movielens}.
\item MovieLens-1M (ML-1M)~\cite{harper2015movielens}.
\item MovieLens Tag Genome 2021 (Genome 2021)~\cite{kotkov2021tag_genome_2021}.
\item Netflix Prize (Netflix)~\cite{bennett2007netflix}.
\item Douban Movie (Douban)~\cite{wang2019huerec}.
\item Amazon Reviews 2018 (Amazon)~\cite{ni2018amazon_review_data}.
\end{compactitem}

\sstitle{Evaluation Metrics}
We report HR@$K$ and NDCG@$K$ for the target item $i_{\mathrm{tar}}$ on test users $\mathcal{U}_{\mathrm{test}}$, where:
\begin{equation}
    \mathrm{HR}@K=\frac{1}{|\mathcal{U}_{\mathrm{test}}|}\sum_{u\in\mathcal{U}_{\mathrm{test}}}\mathbb{I}[i_{\mathrm{tar}}\in\mathrm{TopK}(u)],
    \label{eq:hr_k}
\end{equation}
\begin{equation}
    \mathrm{NDCG}@K=\frac{1}{|\mathcal{U}_{\mathrm{test}}|}\sum_{u\in\mathcal{U}_{\mathrm{test}}}\frac{\mathbb{I}[i_{\mathrm{tar}}\in\mathrm{TopK}(u)]}{\log_2(\operatorname{rank}(i_{\mathrm{tar}}\mid u,\mathcal{I}\setminus\mathcal{I}_u^+)+1)}
    \label{eq:ndcg_k}
\end{equation}
To verify that AGAS preserves recommendation quality for test users, we report Rec@K on their held-out interactions. For each $u \in \mathcal{U}_{\mathrm{test}}$, we check whether $i_u^{\mathrm{test}}$ appears in the poisoned model's top-$K$ list:
\begin{equation}
    \mathrm{Rec}@K=\frac{1}{|\mathcal{U}_{\mathrm{test}}|}\sum_{u\in\mathcal{U}_{\mathrm{test}}}\mathbb{I}[i_u^{\mathrm{test}}\in\mathrm{TopK}(u)]
    \label{eq:recK}
\end{equation}
A high Rec@K indicates that the poisoned model continues to serve genuine users well. This suggests that AGAS promotes the target item without corrupting the broader recommendation behavior. Stealth is measured by detector performance (Accuracy, Precision, Recall, F1) under representative anomaly checks. Lower scores mean the attack is harder to detect.

\sstitle{Victim Models}
We evaluate across CF recommenders spanning matrix factorization, graph convolution, and contrastive paradigms: MF~\cite{koren2009matrix_factorization}, MF (BPR)~\cite{rendle2009bpr}, NeuMF~\cite{he2017neumf}, GMF~\cite{he2017neumf}, NCF~\cite{he2017neumf}, NGCF~\cite{wang2019ngcf}, LightGCN~\cite{he2020lightgcn}, SimGCL~\cite{yu2022simgcl}, XSimGCL~\cite{yu2023xsimgcl}, EGCF~\cite{zhang2024egcf}, and LightCCF~\cite{zhang2025lightccf}.

\sstitle{Attack Baselines} 
We compare against representative shilling and poisoning baselines spanning heuristic, optimization-based, and agentic methods: NoneAttack, RandomAttack~\cite{omahony2005attack_types}, BandwagonAttack~\cite{omahony2005attack_types}, AUSH~\cite{lin2020aush}, PoisonRec~\cite{fang2020poisonrec}, GSPAttack~\cite{nguyen2023gspattack}, CLeaR~\cite{wang2024unveiling_contrastive_poisoning}, PGA~\cite{li2016pga_cf_poisoning}, TargetedAttack~\cite{guo2023targeted_gnn_shilling}, AgentSA~\cite{gu2026llm_agent_shilling_wsdm}, and AgentAttack~\cite{li2026agentattack}.

\sstitle{Comparison Protocol}
Unless a section explicitly studies budget scaling, all methods are evaluated under the same fake-user budget and target-item set for each dataset--victim pair. For agentic methods, the interaction budget is matched by imposing the same upper limit on injected actions across fake users. After each poisoned dataset is constructed, the victim is retrained using the same retraining protocol.

\sstitle{Feedback Protocol}
AGAS follows an offline-transfer protocol. During attack generation, the Coordinator observes rank movement only through recommendation lists returned to probe workers, typically PR/CA workers that have not rated the target. The evaluated victims are then trained from scratch on the clean matrix and on the poisoned matrix $\mathbf{R}'=[\mathbf{R};\tilde{\mathbf{R}}]$~(\autoref{eq:poison_append}). AGAS never queries evaluated victims using $\mathcal{U}_{\mathrm{test}}$.



\subsection{Performance}
\label{sec:performance}
To address \textbf{RQ1 (Performance)}, we evaluate AGAS against baselines on diverse datasets, victim models, and metrics under the matched protocol of \autoref{sec:experiments}. \autoref{tab:bench_unpop} shows that AGAS achieves the best result in every Tail-regime (Unpopular Items, \autoref{fig:intro_radar_motivation}) setting, with sizable gains over the strongest baseline across matrix factorization, graph convolution, and graph contrastive victim families. It stays effective without attacking aggressively in every round, since it may probe the victim early, slow down as suppression signals increase, or pause when the campaign becomes risky. Such non-myopic behavior is beneficial: temporarily reducing intensity preserves workers, avoids unstable promotion patterns, and produces more reliable long-term target movement. On the Head/Mid regimes (Popular Items, \autoref{fig:intro_radar_motivation}), \autoref{fig:popularity_regime} shows AGAS dominant on every axis. AgentAttack and AgentSA are the closest baselines, but the gap remains consistent across victim types and popularity regimes, indicating that AGAS is not tied to one target regime or one victim family.
\subsection{Stealthiness}
\label{sec:stealth}

To address \textbf{RQ2 (Stealthiness)}, we examine whether AGAS promotes targets while preserving normal preference structure. \autoref{fig:stealth_tsne_combined} reports two t-SNE views: the left measures \emph{per-user preference preservation}, where the target should move closer to the user without pulling the user away from ground-truth items; the right measures \emph{global distributional camouflage}, where fake histories should overlap with benign users rather than form separate clusters. AGAS is the only method that keeps both target and ground-truth distances small, showing that it promotes the target while preserving the user's original preference region. It also spreads fake histories more diffusely within the benign embedding cloud, whereas other methods form more concentrated fake clusters. This stealth behavior comes from AGAS's round-level adaptation and role-switching, which prevent fake users from collapsing into repetitive attack patterns.


\begin{figure}[t]
    \centering
    \includegraphics[width=0.85\linewidth]{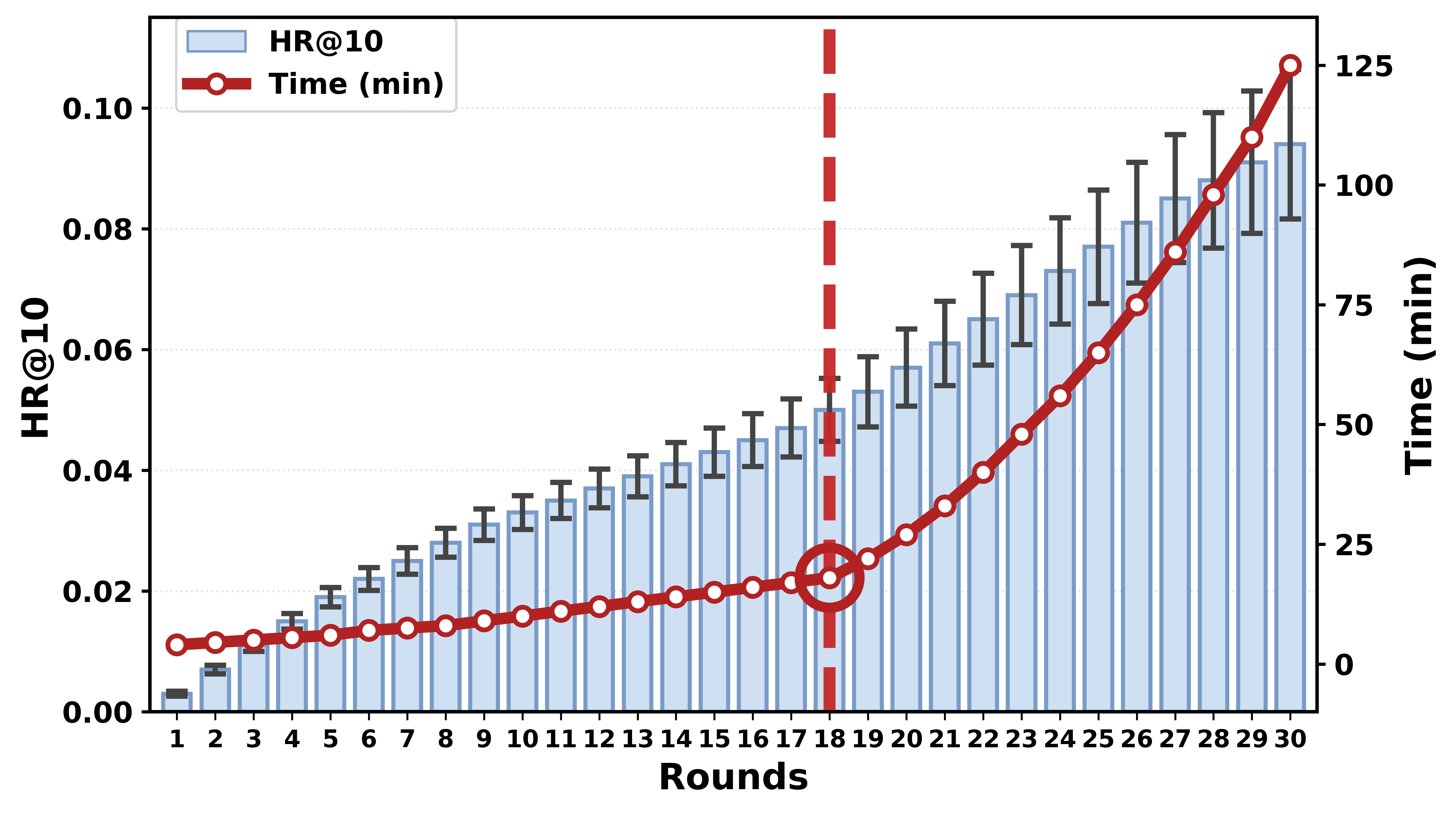}
    \caption{Round-wise trade-off between HR@10 and runtime (minutes). Error bars show 95\% CI.}
    \vspace{-.5em}
    \label{fig:efficiency_rounds_tradeoff}
\end{figure}

\begin{figure}[t]
    \centering
    \includegraphics[width=0.85\linewidth]{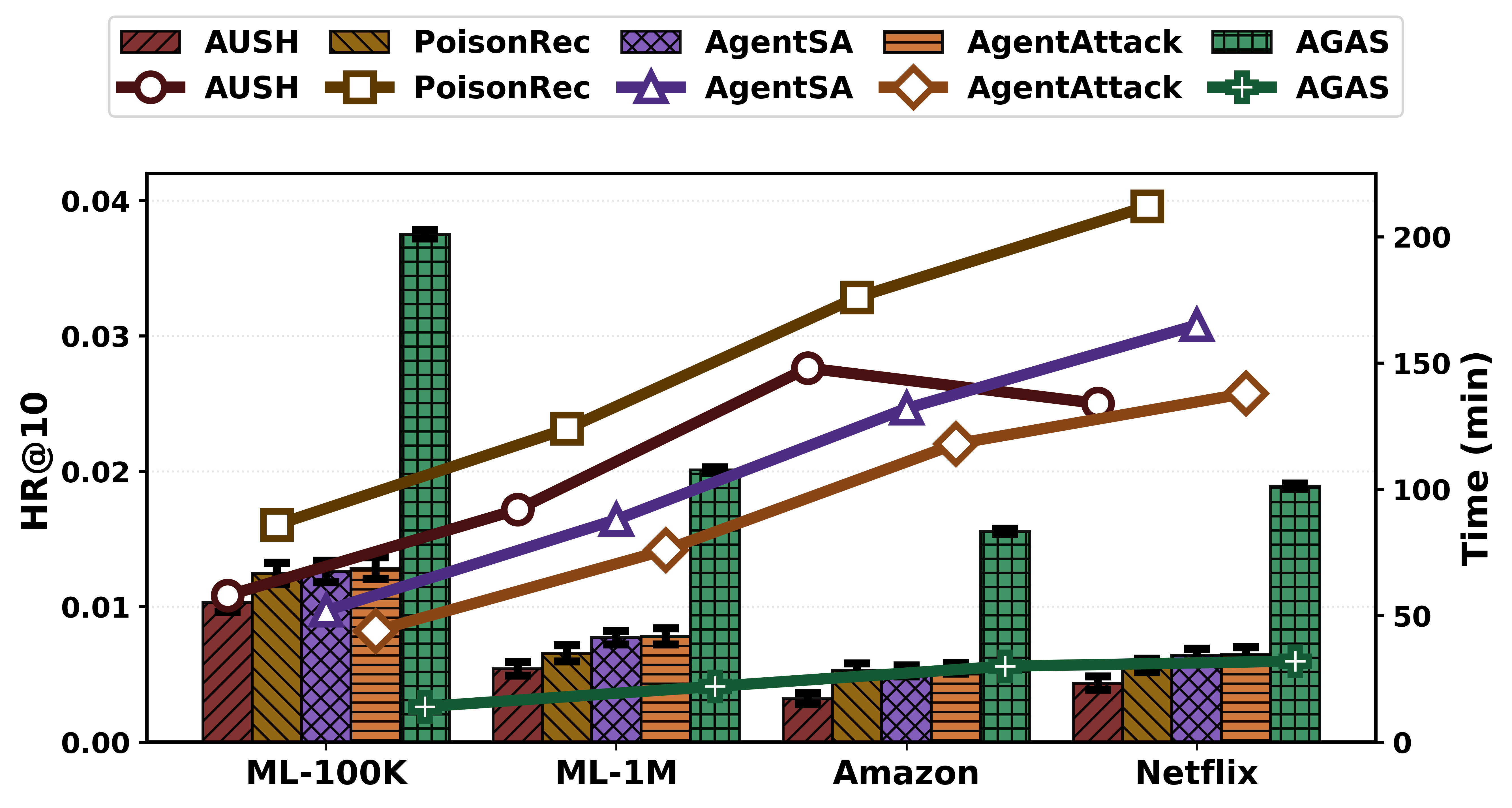}
    \caption{Bars show HR@10 on the left axis, and lines show total runtime (minutes) on the right axis.}
    \vspace{-.5em}
    \label{fig:efficiency_method_time_compare}
\end{figure}

\begin{figure}[t]
    \centering
    \includegraphics[width=0.9\linewidth]{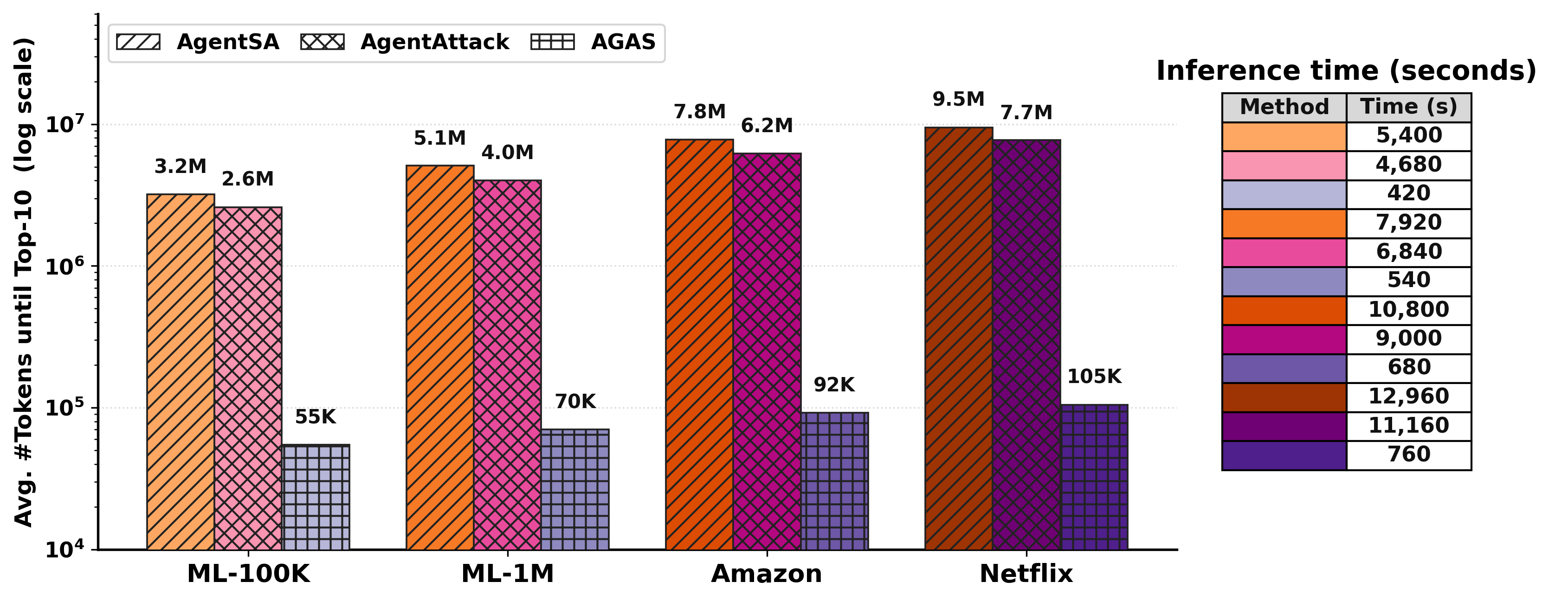}
    \caption{\emph{Left:} bar chart (log scale) of the average number of LLM tokens until the target first enters the Top-10. \emph{Right:} the corresponding wall-clock seconds for each dataset.}
    \vspace{-.5em}
    \label{fig:efficiency_token_topk}
\end{figure}

\begin{figure}[t]
    \centering
    \includegraphics[width=0.7\linewidth]{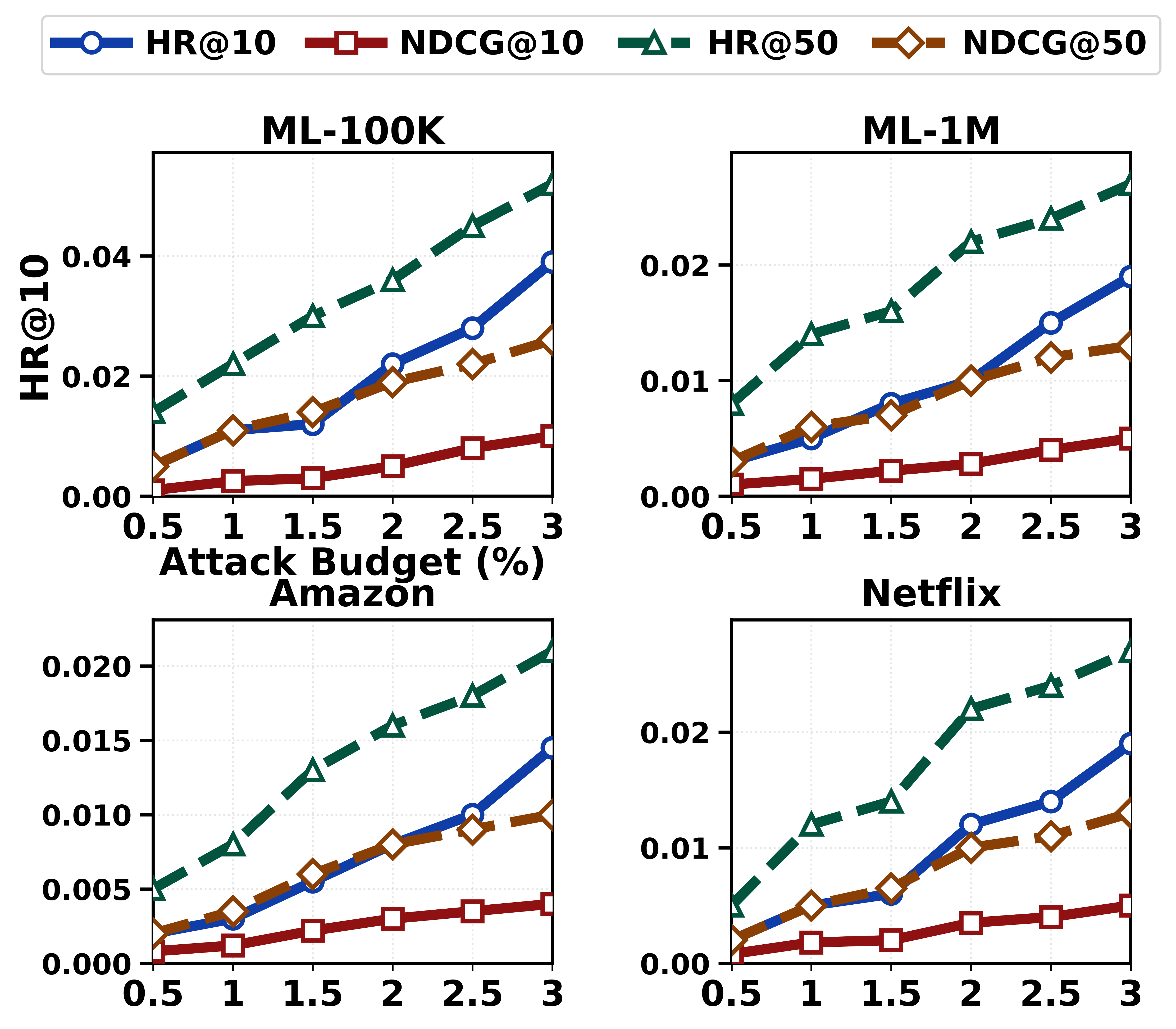}
    \caption{Each panel reports performance as the injected fake-user budget grows: $0.5$--$3.0\%$ of the user population.}
    \vspace{-.5em}
    \label{fig:efficiency_budget_tradeoff}
\end{figure}

\subsection{Detector}
\label{sec:detector}

To address \textbf{RQ3 (Detector)}, we evaluate AGAS against five representative shilling detectors: \textbf{BaseDetect}~\cite{williams2007defending}, \textbf{DHAGCN}~\cite{hao2023detection}, \textbf{PCASelectUsers}~\cite{mehta2009unsupervised}, and two group-level detectors, \textbf{GAGE}~\cite{zhang2020gage} and \textbf{MD-CBA}~\cite{xu2023group_shilling}. \autoref{tab:detect_mf} reports Accuracy, Recall, Precision, and F1, where \emph{lower} values mean the attack is harder to detect. Non-group detectors identify PoisonRec and AgentSA through profile statistics, graph neighborhoods, and PCA-aligned outliers, while group detectors capture cross-user overlap in AgentSA's prompt-driven histories. AgentAttack reduces duplicate-like behavior by mixing several attack families, but pre-trained templates still leave residual group signatures. AGAS instead lowers all metrics across all five detectors by distributing payload actions across roles and rounds. PR and CA build benign-looking context anchored to receptive benign users (\autoref{sec:stealth}), SN is used only by cleaner workers, and IN breaks synchronized activity, so both per-profile outliers and group-overlap signatures become weaker. \autoref{fig:detect_rec10_victim} further shows that AGAS stays closest to the NoneAttack Rec@10 ceiling across all victims, preserving benign Top-10 quality while promoting the target. Together with \autoref{sec:performance}, these results suggest that current detectors struggle with adaptive agentic campaigns.
\subsection{Ablation}
\label{sec:ablation}

To address \textbf{RQ4 (Ablation)}, we ablate AGAS at the strategy and backbone levels. In \autoref{fig:ablation_strategy_heatmap}, removing \emph{Silent Slowdown}, \emph{Profile Cleanup}, \emph{Safe Replacement}, or \emph{Main Attack} causes the largest drops, so AGAS needs both strong promotion and risk control. \emph{Bridge Building} is graph-specific, hurting LightGCN and LightCCF more than NeuMF because it supports two-hop attack paths. \autoref{tab:ablation_backbones} shows that stronger backbones improve the attack, yet AGAS stays best across all model blocks and metrics and degrades gracefully with smaller ones. Removing the \emph{Coordinator}, \emph{Memory}, or \emph{Signals} is sharpest on Gemma 2 2B (\autoref{fig:ablation_size_heatmap}): for a small language model with small reasoning capacity, AGAS may have to lean more heavily on coordination, persistent memory, and explicit feedback.
\subsection{Efficiency}
\label{sec:efficiency}

To address \textbf{RQ5 (Efficiency)}, we evaluate round-wise trade-offs, runtime, and fake-user budget scaling. HR@10 keeps rising with more rounds (\autoref{fig:efficiency_rounds_tradeoff}). We set 18 rounds as the AGAS budget because later rounds become increasingly costly, adding wall-clock time and slightly wider 95\% confidence intervals. This horizon also gives the Coordinator time to adjust roles before stronger promotion, making AGAS appear more normal (\autoref{sec:stealth}) and harder to detect (\autoref{sec:detector}). \autoref{fig:efficiency_method_time_compare} and \autoref{fig:efficiency_token_topk} show that AGAS reaches the highest HR@10 at modest cost, about 24 minutes and tens of thousands of tokens to push the target into the Top-10, whereas AgentAttack needs millions of tokens and other methods hours of training. AGAS fits no GAN, solves no bi-level objective, retrains no per-target surrogate, and pre-generates no profile library. Its cost is one Coordinator decision plus short worker calls per round, so runtime and token use grow approximately linearly with rounds and fake users. Gains stay steady as the budget grows from $0.5\%$ to $3.0\%$, with smaller gains on Amazon indicating that harder datasets require more attack pressure (\autoref{fig:efficiency_budget_tradeoff}). AGAS is thus \emph{budget-aware}: rounds, tokens, and fake-user portion can be tuned to trade cost for HR@10, making it a \emph{cost-aware benchmark framework} for RecSys robustness.

\section{Conclusion}
\label{sec:conclusion}
This paper introduced AGAS, a black-box agentic framework that orchestrates coordinated group shilling attacks through dynamic role switching and multi-stage campaigns. The Coordinator manages attack escalation, slowdown, pausing, replacement, and round-level adaptation across a group of fake-user workers. Across diverse datasets and CF-based victim models, AGAS achieves stronger target promotion than prior baselines while preserving more realistic user behavior and benign recommendation quality under matched evaluation budgets. It runs end-to-end without costly fine-tuning or repeated surrogate retraining, reaches strong attacks at modest runtime and token cost, and transfers across victim families, making it both a practical threat model and a reusable benchmark for stress-testing recommender robustness. Future work should extend AGAS to more complex victims such as multimodal recommenders, or LLM-empowered recommenders. As Agentic Web capabilities spread, recommender systems must defend against attacks that adapt to their defenses at scale.

\section*{Acknowledgment}
This work was partially supported by the Australian Research Council through the Discovery Project (Grant Nos. DP260100326 and DP240101108), the Linkage Projects (Grant Nos. LP230200892 and LP250200778), and the DECRA Project (Grant No. DE260100673).



\bibliographystyle{IEEEtran}
\bibliography{IEEEabrv,ref}

\end{document}